\documentclass[aps,prd,twocolumn,amsmath,amssymb,amsfonts,nofootinbib,superscriptaddress,altaffilletter]{revtex4-1}

\usepackage{graphicx}% Include figure files
\usepackage{dcolumn}% Align table columns on decimal point
\usepackage{bm}% bold math
\usepackage{url}
\usepackage{subfigure}
\usepackage{float}
\usepackage{amssymb}
\usepackage{amsmath}
\usepackage{longtable}
\usepackage{rotating}
\usepackage{color}
\usepackage{xcolor}
\usepackage{epsfig}
\usepackage{epsf}
\usepackage{fancyhdr}
\usepackage{hyperref}
\newcommand{\asini}{a_p}
\newcommand{\asinisour}{a_{p_{s}}}
\newcommand{\argp}{\omega}
\newcommand{\argpsour}{\omega_s}
\newcommand{\ecc}{e}
\newcommand{\tPeri}{t_p}
\newcommand{\tPerisour}{t_{p_{s}}}
\newcommand{\tEM}{\tau}
\newcommand{\Om}{\Omega}
\newcommand{\tAsc}{t_\mathrm{asc}}
\newcommand{\Tobs}{T_\mathrm{obs}}
\newcommand{\Tfft}{T_\mathrm{FFT}}

\newcommand{\fdot}{\dot{f}}
\newcommand{\Rth}{\mathcal{R}_{\mathrm{th}}}
\newcommand{\atmid}{{\mathrm{mid}}}
\newcommand{\tMid}{t_\atmid}
\newcommand{\tref}{t_\mathrm{ref}}

\newcommand{\tArr}{t_{\mathrm{arr}}}
\newcommand{\tSSB}{t_{\mathrm{SSB}}}

\newcommand{\highlight}[1]{%
  \colorbox{gray!90}{$\displaystyle#1$}}
\newcommand{\LightHighlight}[1]{%
  \colorbox{gray!54}{$\displaystyle#1$}}
\newcommand{\lowHighlight}[1]{%
  \colorbox{gray!30}{$\displaystyle#1$}}
 \newcommand{\cosi}{\cos\iota}
 \newcommand{\cosisour}{\cos \iota_{s}}

\begin{document}

\title{Novel directed search strategy to detect continuous gravitational waves from neutron stars in low- and high-eccentricity binary systems} 

\author{Paola~Leaci}
\altaffiliation{paola.leaci@roma1.infn.it}
\affiliation{Dip. di Fisica, Universit\`{a} di Roma ``Sapienza'', P.le A. Moro, 2, I-00185 Rome, Italy}
\affiliation{INFN, Sezione di Roma, P.le A. Moro, 2, I-00185 Rome, Italy}
\author{Pia~Astone}
\affiliation{INFN, Sezione di Roma, P.le A. Moro, 2, I-00185 Rome, Italy}
\author{Sabrina~D'Antonio}
\affiliation{INFN, Sezione di Roma 2, Via della Ricerca Scientifica, 1, I-00133 Roma, Italy}
\author{Sergio~Frasca}
\affiliation{Dip. di Fisica, Universit\`{a} di Roma ``Sapienza'', P.le A. Moro, 2, I-00185 Rome, Italy}
\affiliation{INFN, Sezione di Roma, P.le A. Moro, 2, I-00185 Rome, Italy}
\author{Cristiano~Palomba}
\affiliation{INFN, Sezione di Roma, P.le A. Moro, 2, I-00185 Rome, Italy}
\author{Ornella~Piccinni}
\affiliation{Dip. di Fisica, Universit\`{a} di Roma ``Sapienza'', P.le A. Moro, 2, I-00185 Rome, Italy}
\affiliation{INFN, Sezione di Roma, P.le A. Moro, 2, I-00185 Rome, Italy}
\author{Simone~Mastrogiovanni}
\affiliation{Dip. di Fisica, Universit\`{a} di Roma ``Sapienza'', P.le A. Moro, 2, I-00185 Rome, Italy}
\affiliation{INFN, Sezione di Roma, P.le A. Moro, 2, I-00185 Rome, Italy}
\date{\today}

\begin{abstract}

We describe a novel, very fast and robust, directed search incoherent method for periodic gravitational waves from neutron stars in binary systems.
As directed search, we assume the source sky position to be known with enough accuracy, but all other parameters (including orbital ones)
are supposed to be unknown. We exploit the frequency-modulation due to source orbital motion to unveil the signal signature by commencing from a
collection of time and frequency peaks (the so-called $\textit{peakmap}$).

We validate our pipeline adding 131 artificial CW signals from pulsars in binary systems to simulated detector Gaussian noise, characterised by a power spectral density $S_h = 4 \times 10^{-24}$~Hz$^{-1/2}$ in the frequency interval $[70,\, 200]$~Hz, which is overall commensurate with the advanced detector design sensitivities. The pipeline detected 128 signals, and the weakest signal injected and detected has a gravitational-wave strain amplitude of $\sim  10^{-24}$, assuming one month of gapeless data collected by a single advanced detector. We also provide sensitivity estimations, which show that, for a single-detector data covering one month of observation time, depending on the source orbital Doppler modulation, we can detect signals with an amplitude of $\sim 7\times 10^{-25}$. By using three detectors, and one year of data, we would easily gain more than a factor 3 in sensitivity, translating into being able to detect weaker signals. 

We also discuss the parameter estimate proficiency of our method, as well as computational budget, which is extremely cheap. In fact, sifting one month of single-detector data and 131~Hz-wide frequency range takes roughly 2.4 CPU hours. Due to the high computational speed, the current procedure can be readily applied in ally-sky schemes, sieving in parallel as many sky positions as permitted by the available computational power.

The novel procedure has a sensitivity comparable and slightly higher than other competing pipelines present in literature, but is several orders of magnitude faster than those.

We also introduce (ongoing and future) approaches to attain sensitivity improvements and better accuracy on parameter estimates in view of the use on real advanced detector data.

\end{abstract}

\maketitle

%%%%%%%%%%%%%%%%%%%
\section{\label{Intro} Introduction}
%%%%%%%%%%%%%%%%%%%
Since the early 1960s, when the first gravitational-wave bar detector was developed~\cite{Weber1960}, the experiments that aimed at the detection of gravitational radiation, planned in laboratories throughout the world, have been in continuous progress~\cite{Auriga,Nautilus,Explorer,Allegro,Bonaldi,LeaciPRA,LeaciPRD,LeaciCQG,Virgo,Geo,TAMA,ligoref,IPWRA}. 
At present, LIGO~\cite{ADvLIGO} and Virgo~\cite{AdvVirgoRef:2009} are the most sensitive ground-based gravitational-wave detectors.

Following a major upgrade lasted for 5 years, with consequent improvement in sensitivity~\cite{AdvDetO1}, the LIGO observatories resumed data taking with the first observing science run during which they have collected data between September 2015 and January 2016. On September 14, 2015, the advanced LIGO interferometers detected for the first time a coincident transient gravitational-wave signal produced by the coalescence of a pair of black holes~\cite{GWevent}, marking thus the official beginning of a new era: the era of gravitational-wave astronomy.

There are however other classes of gravitational-wave signals, which have still to be detected, such as long-lived continuous waves (CWs), which are expected to be emitted by rapidly rotating neutron stars (NSs) with nonaxisymmetric deformations~\cite{Owen:2005fn}. We expect $\mathcal{O}(10^8)$ of these sources to exist in the Galaxy, but only $\sim2\,500$ NSs (mostly pulsars) have been electromagnetically observed~\cite{atnf}. Roughly 1\,300 of these observed radio pulsars are located in binary systems, and have rotation rates that can allegedly emit CWs in the advanced LIGO-Virgo sensitivity band. This promising class of sources is the target of the current paper. 

The detection of CW signals will enrich the understanding we have about the emitting objects (i.e., NSs), providing us insight about the equation of state of the matter at supranuclear densities inside the NSs, know exactly the NS degree of asymmetry, and have also demographic and evolutionary information about these sources.

In the rest frame of the NS, CWs have a constant amplitude and are quasimonochromatic with a slowly decreasing intrinsic frequency. They are received at Earth-based detectors with a Doppler modulation due to the relative motion between the source and the detector. Consequently, the observed phase evolution depends on the intrinsic signal frequency, frequency time derivatives (also known as ``spindown'' terms), and source sky position. If the source is located in a binary system, there is a further frequency-modulation caused by the source orbital motion, which in general is described by five unknown Keplerian parameters~\cite{Dhurandhar:2000sd}, as detailed in Sec.~\ref{sec:binary-cw-phase}.

The weakness of the expected signal requires long integration times, typically of the order of a few months or years, to accumulate a signal-to-noise ratio (SNR) sufficient for detection as, for a coherent (incoherent) search, the SNR scales as the square (fourth) root of the length of the observational time (i.e. the length of the data being analyzed)~\cite{Jaranowski:1998qm,Astone:2000jza}.
All-sky, wide frequency searches over long observation times cannot be treated by using standard coherent methods (where the phase information is used), as is the case for targeted and narrow-band searches for known pulsars~\cite{VelaSpinDAbadie:2011md,NarrowBandVelaCrabAasi:2014jln}, because of the demanding computational burden. Hence, hierarchical approaches have been proposed~\cite{Brady:1998njStackSlide,Krishnan:2004sv,Cutler:2005pn}, where the entire data set is split into shorter segments. Each segment is analyzed coherently, and afterwards the information from the different segments is combined incoherently (which means that the phase information is lost). The hierarchical approaches allow us to dramatically reduce the analysis computational time at the cost of a relatively small sensitivity loss.

The additional source orbital parameters make the sieved parameter space to blow up, resulting in a prohibitive computational cost.
Hence, it becomes pressing to develop robust strategies to detect CWs emitted by NSs orbiting a companion object, and being able to reach a tradeoff between computational cost and sensitivity. 
Although CWs have not been detected so far by analysing data from initial LIGO and Virgo detectors, stringent upper limits have been set on the gravitational-wave signal strength for both isolated pulsars~\cite{Aasi:2012fw,Aasi:2015rar,GalcticCAasi:2013jya,NarrowBandVelaCrabAasi:2014jln} and pulsars in binary systems~\cite{TwoSpectAllSkyAasi:2014erp,ScoX1directAasi:2014qak}.

A particularly interesting type of potential CW sources are NSs in low-mass x-ray binaries (LMXBs), with Scorpius X-1 being its most prominent representative~\cite{Watts:2008qw}.
Several searches for CW signals from Scorpius X-1 have been performed (without any detections) on data from initial LIGO~\cite{Abbott:2006vg,ScoX1directAasi:2014qak}, and new pipelines have been developed~\cite{Leaci:2015bka} and recently tested in a Scorpius X-1 mock data challenge (MDC)~\cite{Messenger:2015tga}.

We present here an incoherent strategy that allows us to perform directed searches for CWs emitted by NSs in binary systems (LMXB like sources), exploring a wide frequency range and source orbital parameters at a paltry computing cost.
We also include investigations of pulsars in eccentric orbits, which we know to exist~\cite{HulseTay,EccOrbitWatts:2016uzu,atnf}.
The method is based on selecting significant peaks from a short Fast-Fourier-Transform (FFT) database (SFDB), and exploiting the frequency modulation pattern produced by the source orbital motion to detect a potential CW signal. We show the performance of the current method to detect CW signals by analysing pure Gaussian noise data to which we add hundreds of fake signals. We consider one month of gapless data, i.e., data taken continuously during the assumed observation time from the Virgo (or equivalently LIGO) detector in its advanced configuration.

We restrict our investigation to constant-frequency CW signals (i.e. we neglect frequency time derivatives). This is motivated by the assumed steady-state torque-balance situation in LMXB like sources, which are our main target of interest. However, the corresponding fluctuations in the accretion rate are expected to cause some stochastic frequency drift, and one will therefore need to be careful to restrict the maximal \textit{coherence time} (i.e. FFT duration) in order to limit the frequency resolution. In fact, in~\cite{Leaci:2015bka} it has been shown that the maximal FFT duration would be restricted by the astrophysical concern of \textit{spin wandering}, namely a stochastic variability of the spin frequency due to variations in the accretion rate. In the present work we neglect the spin wandering effect, as our new robust methodology is expected to be unaffected by these variations. Exhaustive future studies will be able to shed light on these considerations and will be presented in a subsequent paper.

The novel procedure we present allows us to detect gravitational-wave signals with strain amplitude of $\sim  10^{-24}$, the weakest value we used for the simulations, assuming one month of gapeless data collected by a single advanced detector. Sensitivity estimation studies (see Sec.~\ref{Sec:SensEstim}) show however that, depending on the source orbital Doppler modulation, the current method can detect signals with an amplitude of $\sim 7\times 10^{-25}$. By using three advanced detectors, and one year of data, we would be able to further improve the sensitivity by a factor greater than 3.

The paper is organized as follows. Section~\ref{Sec:signal} provides a general and brief introduction of the expected signal model. Section~\ref{Sec:AnM} describes details on data analysis approach necessary to produce input data set. In Secs.~\ref{Sec:PS} and~\ref{FFTfracOrbit} we discuss the choices of the investigated parameter space and FFT duration. In Sec.~\ref{Sec:method} we present a rigorous description of the novel strategy used to detect CW signals orbiting a companion object. Sections~\ref{RecoveryEstim} and~\ref{Sec:h0recovery} show the key results of the procedure. Sensitivity estimates and computing cost budget are detailed in Secs.~\ref{Sec:SensEstim} and~\ref{Sec:TeoCtot}, respectively. Finally, Sec.~\ref{Sec:Conc} contains concluding remarks, underway search improvements, and future prospects.

%*******************************************
\section{\label{Sec:signal} The signal}
%*******************************************
In the following we briefly recall the expected waveform model and signal phase. 

%****************************************************************************************************
\subsection{\label{Sec:h0} The waveform model}
%****************************************************************************************************
The expected waveform of a nonaxisymmetric NS, rapidly rotating around one of its principal axes, and received at the detector is

\begin{align}
\label{Eq:hSignal}
h(t)  = h_0\,F_+(t;\vec{n},\psi)\,  \frac{1+\cos^2\iota}{2}\, \cos \phi(t) +\nonumber\\
h_0\,F_\times(t;\vec{n},\psi)\, \cos\iota\,\sin \phi(t), 
\end{align}
where $\iota$ is the inclination angle of the NS rotational axis to the line of sight, $F_{+,\times}$ are the detector beam pattern functions to \textit{plus} and \textit{cross} polarized gravitational waves, which depend on the sky position $\vec{n}$ and the relative polarization angle $\psi$ of the wave-frame\footnote{The wave-frame is a right-handed Cartesian coordinate system based on the direction of propagation of the gravitational wave. Its z-axis is along the direction of propagation, and its x-and y-axes are along the principal directions of polarization of the wave.} \cite{Jaranowski:1998qm,Bonazzola:1995rb}. In standard equatorial coordinates with right ascension $\alpha$ and declination $\delta$, the components of the unit vector $\vec{n}$, pointing to the NS, are given by $(\mathrm{cos}\,\alpha \, \mathrm{cos}\,\delta,\ \mathrm{sin}\,\alpha \, \mathrm{cos}\,\delta,\ \mathrm{sin} \,\delta)$. The phase evolution $\phi(t)$ of the gravitational-wave signal is described in Sec.~\ref{sec:binary-cw-phase}.  

The gravitational-wave amplitude parameter is given by
\begin{equation}
\label{eq:GWampl}
  h_0 = \frac{4\pi^2G}{c^4}\frac{I_{zz}f^2\varepsilon}{d},
\end{equation}
where $f$ is the frequency of the emitted CW signal (which is also twice the rotational frequency of the star for the sources we are interested in~\cite{IPWRA}); $G$ is Newton's constant, $c$ is the speed of light,  $d$ is the distance to the star, and  $\varepsilon$ is the star ellipticity expressed in terms of principal moments of inertia. 
The distribution of $\varepsilon$ for NSs is uncertain and model dependent since the breaking strain for a NS crust is highly uncertain (see, e.g.,~\cite{NonAxNS2,Smoluchowski,Ruderman1991,Horowitz} for exhaustive discussion).

Spinning NSs in binary systems are particularly interesting because accretion from a companion may cause an asymmetrical quadrupole moment of inertia of the spinning NS. An intriguing astrophysical model postulating torque balance between the accretion and CW emission~\cite{papaloizou78:_gravit,wagoner84:_gravit,bildsten98:_gravit} yields a predicted CW amplitude that, for Scorpius~X-1, is
\begin{equation}
  \label{eq:71}
  h_0 \sim 3.5\times10^{-26} \sqrt{\frac{300\, \mathrm{Hz}}{f}}\,.
\end{equation}

%****************************************************************************************************
\subsection{Binary CW signal phase}
\label{sec:binary-cw-phase}
%****************************************************************************************************
The general CW phase model assumes a slowly spinning-down NS with a rotation rate and quadrupolar deformation resulting in the emission of CWs. The phase evolution can therefore
be expressed as a Taylor series in the NS source frame as
\begin{equation}
  \label{eq:12}
  \phi^{\mathrm{src}}(\tau) = 2\pi\,[ f\,(\tau-\tref) +
    \frac{1}{2}\dot{f}(\tau-\tref)^2 + \ldots ]\,,
\end{equation}
where $\tref$ denotes the reference time and $f,\dot{f},\ddot{f},\ldots$ are the CW frequency and spindown parameters. 

To relate the CW phase in the source frame to the phase $\phi(\tArr)$ in the detector frame, we need to relate the wavefront detector arrival time $\tArr$ to its
source emission time $\tau$, i.e.\ $\tau(\tArr)$, such that $\phi(\tArr) = \phi^{\mathrm{src}}(\tau(\tArr))$.
Neglecting relativistic wave-propagation effects, such as Einstein and Shapiro delays (see, e.g., \cite{2006MNRAS.369..655H,2006MNRAS.372.1549E} for
more details), we can write this as
\begin{equation}
  \label{eq:13}
  \tau(\tArr) = \tArr + \frac{\vec{r}(\tArr)\cdot \vec{n}}{c} - \frac{D}{c} - \frac{R(\tau)}{c}\,,
\end{equation}
where $\vec{r}$ is the vector from solar-system barycenter (SSB) to the detector, $D$ is the (generally unknown) distance between the SSB and the binary barycenter (BB), $R$ is the radial distance of the CW-emitting NS from the BB along the line of sight, where $R > 0$ means the NS is further away from us than the BB.

Following the discussion in Sec.~III~A of~\cite{Leaci:2015bka}, we can write the R\o{}mer delay of the binary (i.e., the light-travel time across the orbit) as
\begin{equation}
  \label{RoemD}
  \frac{R}{c} = \asini\left[ \sin\,\argp ( \cos E - \ecc) + \cos\argp\,\sin E \sqrt{1 - \ecc^2} \right]\,,
\end{equation}
where we defined the projected semimajor axis $\asini\equiv a\sin{I}/c$ of the NS orbit; $I$ is the inclination angle between the orbital plane and the sky, $a$ the semi-major axis, $\argp$ is the argument of periapse, $\ecc$ the orbital eccentricity, and $E$ the \emph{eccentric anomaly}, defined by the transcendental relation
\begin{equation}
  \label{eq:KeplEq}
  \tEM - \tPeri = \frac{P}{2\pi}\left( E - \ecc\,\sin E\right)\,.
\end{equation}
Equation~\eqref{eq:KeplEq} is the so-called Kepler's equation, and describes the dynamics in binary systems; $P$ 
is the binary period, and $ \tPeri$ is the time of periapse passage.

Dropping the unknown distance $D$ to the BB (which is equivalent to re-define the intrinsic spindown parameters), and defining the SSB wavefront arrival time $\tSSB$ as
\begin{equation}
  \label{eq:19}
  \tSSB(\tArr;\vec{n}) \equiv \tArr + \frac{\vec{r}(\tArr)\cdot\vec{n}}{c}\,,
\end{equation}
we can rewrite the timing relation Eq.~\eqref{eq:13} as
\begin{equation}
  \label{eq:20}
  \tau(\tSSB) = \tSSB - \frac{R(\tau)}{c}\,.
\end{equation}
As we are interested only in binary systems with known sky-position $\vec{n}$, we can place us into the SSB, which is always possible for known $\vec{n}$~\cite{Leaci:2015bka}. In order to simplify the notation, we now simply write $t \equiv \tSSB$.
Plugging this timing model into the phase of Eq.~\eqref{eq:12}, we obtain
\begin{equation}
  \label{eq:21}
  \phi(t) \approx 2\pi\left[ f \left(\Delta t - \frac{R}{c}\right) + \frac{1}{2}\fdot \left(\Delta t - \frac{R}{c}\right)^2 + \ldots\right]\,,
\end{equation}
with $\Delta t \equiv t - \tref$. The binary systems we are interested in have semi-major axis $\asini$ of order of $\mathcal{O}(1-3)\,$s, and binary periods $P$ of order of several hours. Hence, the change in $E$, and therefore $R(E)$ during the time $R/c$, will be negligible, and so we can approximate
$E(\tau) \approx E(t)$, namely
\begin{equation}
  \label{eq:22}
  t - \tPeri \approx \frac{P}{2\pi}\left( E - \ecc\,\sin E\right)\,.
\end{equation}
For our purposes, by using the linear phase-model approximation, we can write Eq.~\eqref{eq:21} as~\cite{Leaci:2015bka}

\begin{equation}
\label{eq:ComplPhase}
\phi(t) \approx 2\pi f \left[\Delta t - \frac{R(t)}{c}\right]  \,.
\end{equation}
Replacing Eq.~\eqref{RoemD} into Eq.~\eqref{eq:ComplPhase}, we find 

\begin{align}
\label{eq:ComplLinPhaseApprox}
\phi(t) = 2\pi f \left\{ \Delta t - \asini \left[ \sin\,\argp ( \cos E(t) - \ecc) +  \right. \right. \nonumber\\
 \left. \left. \cos\argp\,\sin E(t) \sqrt{1 - \ecc^2} \right] \right\}\,,
\end{align}
which is the phase model valid for eccentric orbits. It is useful to consider, however, also the \emph{small-eccentricity limit}, thus simplifying Eqs.~\eqref{RoemD} and~\eqref{eq:ComplLinPhaseApprox}. To this purpose, we Taylor expand Eqs.~\eqref{RoemD}~and~\eqref{eq:22} up to leading order in $\ecc$, i.e., inserting
$E(t) = E_0(t) + \ecc E_1(t) + \ldots$ into Kepler's equation [i.e., Eq.~\eqref{eq:22}], and obtain
\begin{align}
  \label{eq:24}
  E_0(t) &= \Om ( t - \tPeri )\,,\\
  E_1(t) &= \sin E_0(t)\,.
\end{align}
Plugging this information into Eq.~\eqref{RoemD}, we obtain the R\o{}mer delay of the binary to leading order in
$\ecc$ as
\begin{equation}
  \label{eq:RoemDsmallEcc}
  \frac{R(t)}{c} = \asini \left[ \sin\psi(t) + \frac{\kappa}{2}\sin2\psi(t) - \frac{\eta}{2}\cos2\psi(t) -\,\frac{3\,\eta}{2}\right]\,.
\end{equation}
We use the standard \emph{Laplace-Lagrange} parameters defined as
\begin{align}
   \kappa &\equiv \ecc\,\cos(\argp)\,, \label{eq:kappa}\\   
  \eta  &\equiv \ecc\,\sin(\argp)\, \label{eq:eta} ,
\end{align}
and the mean orbital phase
\begin{equation}
  \label{eq:27}
  \psi(t) \equiv \Om\,(t - \tAsc)\,,
\end{equation}
measured from the \emph{time of ascending node} $\tAsc$, which (for small $\ecc$) is related to $\tPeri$ by \cite{2006MNRAS.369..655H}
\begin{equation}
  \label{eq:28}
  \tAsc \equiv \tPeri - \frac{\argp}{\Om}\,,
\end{equation}
and that (contrary to the time of periapse $\tPeri$), remains well-defined even in the limit of
circular orbits ($\ecc = 0$). The parameter $\Om \equiv \frac{2\pi}{P}$ is the \emph{mean} orbital angular velocity.

The small-eccentricity phase model can be therefore written as

\begin{align}
  \label{eq:EccAnPhase}
\phi(t)\approx 2\pi f \left\{ \Delta t - \asini \left[ \sin\psi + \frac{\kappa}{2}\sin2\psi - \frac{\eta}{2}\cos2\psi  \right. \right. \nonumber\\
 \left. \left. \,-\frac{3\,\eta}{2}  \right] \right\}\,.
\end{align}

%************************************************************
\section{\label{Sec:AnM} Data Analysis Background}
%************************************************************

In what follows we summarise details of the SFDB and \textit{peakmaps} construction, starting from short FFTs of the calibrated detector strain data. 

%****************************************************************************************************
\subsection{\label{Sec:SFDB} Short Fast-Fourier-Transform database}
%****************************************************************************************************
The first step of the analysis consists of building up a short FFT database, the SFDB~\cite{Astone:2005fj}, where the duration of each FFT, i.e. the \textit{coherence time} $\Tfft$, must be short enough such that the signal power remains confined within a frequency bin. The signal frequency changes in time, however, due to the Earth Doppler modulation, source spindown (if present), and -if the star is in a binary system- also due to the modulation caused by the source binary orbit. In Sec.~\ref{FFTfracOrbit} we describe how the FFT time baseline can be constrained in order to take into account this modulation, which we have seen in Sec.~\ref{sec:binary-cw-phase} to be described by five (generally unknown) Keplerian parameters. We note that the only Earth Doppler modulation would allow us to produce FFTs longer than those used here (see e.g. Table~I of~\cite{Astone:2014esa}).

The FFTs are then obtained from calibrated detector strain data split into interlaced (by half) chunks of  $\Tfft$ duration, each windowed in order to limit the dispersion of power due to their finite length.
A time-domain data cleaning procedure, described in~\cite{Astone:2005fj,Acernese:2009zz, PTDC}, is applied when constructing the SFDB to safely remove time-domain disturbances in detector data, which would enhance the detector noise level at the cost of a reduction in search sensitivity.

%****************************************************************************************************
\subsection{\label{Sec:Peakmap} Peakmap}
%****************************************************************************************************
From the SFDB we create a time-frequency map, called \textit{peakmap}, obtained selecting the most significant peaks on equalized periodograms (according to what described in the following). This is a subtle step as the peak selection will affect the detection efficiency: all potential CW candidates, skipped at this stage due to an inaccurate construction of the peakmap, will be definitely lost. The peakmap has been described in~\cite{Astone:2005fj, Astone:2014esa}, but here we recall only the salient aspects.

The peakmap production begins by computing, for each of the $\mathcal{N}$ FFTs in the SFDB, the ratio $\mathcal{R}$ of the square root of the periodogram [i.e. the square modulus of the $i$th FFT, $S_{P;i}(f)$] and the autoregressive  average spectrum estimation, $S_{\mathrm{AR};i}(f)$: 
\begin{equation}
  \label{RdetStat}
 \mathcal{R}(i,\ell) = \sqrt{\frac{S_{P;i}(f_{\ell})}{S_{\mathrm{AR};i}(f_{\ell})}}; \quad i= 1, \,\dots,\, \mathcal{N}\,,
\end{equation}
where $\mathcal{R}$ is computed for every frequency bin (indexed by $\ell$) of the $i$th FFT. By construction the ratio $\mathcal{R}$ is an adimensional function varying around 1 and showing evident departures from 1 when spectral peaks are present. 

The function $\mathcal{R}$ is compared to a threshold $\Rth=\sqrt{2.5}$~\cite{Astone:2014esa} such that, all frequency bins above $\Rth$, and that are local maxima, are selected. We call \emph{peak} each pair consisting of a selected frequency bin and beginning time of the corresponding FFT. 
In general, the collection of all peaks, selected from all FFTs of the SFDB, forms the peakmap. A collection of peaks selected from a single FFT is instead referred to as \textit{subpeakmap}. %Hence, we can have at most $

\begin{figure}[htbp] 
  \centering
  \includegraphics[width=1.01\columnwidth]{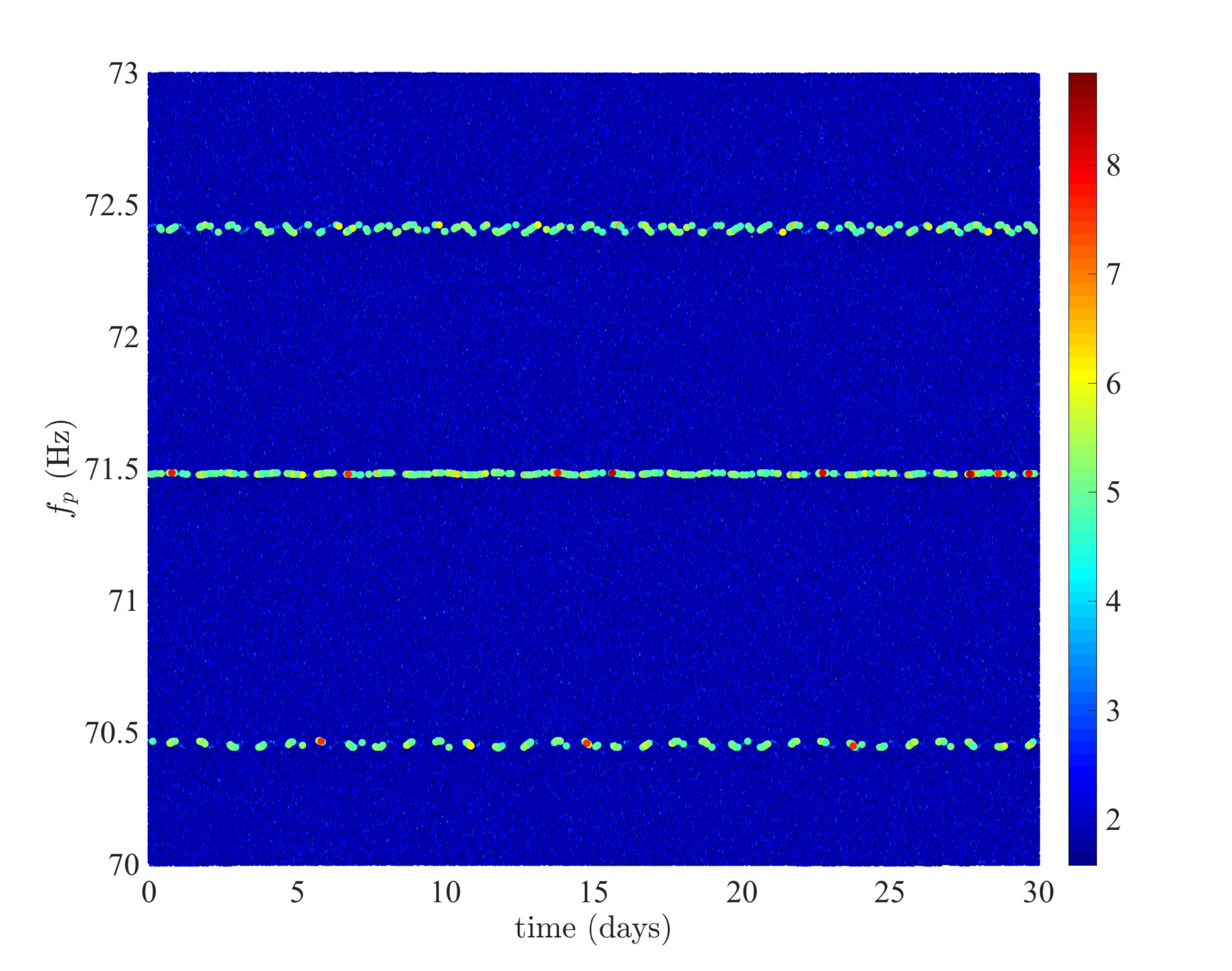} 
  \caption{(Color online) Peakmap in the frequency band [70,\,73]~Hz, where we can recognise the track of three simulated CW signals. Time is since the beginning of the run.} 
  \label{fig:PM7073}
\end{figure}

An example of peakmap (corrected by the Earth's Doppler modulation) is shown in Fig.~\ref{fig:PM7073} for simulated data covering an observation time of $30$~days, and a detector power spectral density $S_h = 4 \times 10^{-24}$~Hz$^{-1/2}$ in the frequency band [70,\,73]~Hz~\footnote{Details on how such fake data have been generated are given in Sec.~\ref{Sec:DataPro}.}. The faint tracks of three (fake) CW signals, having a gravitational-wave strain amplitude of $\sim 3 \times 10^{-24}$, and an adimensional Doppler modulation due to orbital motion of $\Delta M \sim 2 \times 10^{-4},\, 6 \times 10^{-5}, \, 2.6 \times 10^{-4}$ [see Eq.~\eqref{eq:modDepth}] for frequency of $\sim70.5$~Hz, 71.5~Hz, and 72.5~Hz, respectively, are clearly visible as sinusoidal curves. We note that the peakmap is corrected only by the modulations caused by the Earth (orbital and rotational) motions, and not by the binary orbital motion (described by unknown parameters). If this last effect could be removed, the signals would appear as straight lines.

We stress that, on the contrary of what implemented in the \textit{StackSLide}~\cite{Brady:1998njStackSlide} and \textit{Powerflux}~\cite{Abbott:2007tdPowerfl} schemes, the peak amplitude $\mathcal{R}$ is not taken into account in the analysis, but only in the selection step~\footnote{We note that we will use later the peak amplitude to recover the gravitational-wave strain amplitude, as outlined in Sec.~\ref{Sec:h0recovery}.}. The selection of peaks above threshold, and that are also local maxima, translates into a better robustness to spectral disturbances and a significant reduction of the analysis computational burden, as the number of selected peaks is smaller. This comes at the cost of only a small sensitivity loss ($\lesssim \% 10$), as described in~\cite{Astone:2014esa} (where we refer a reader to for statistical considerations about peak selection).

%****************************************************************************************************
\section{\label{Sec:PS} Search parameter space}
%****************************************************************************************************

We describe the steps performed to create a fake data set to which we added simulated CW signals, which have the scope to validate the new algorithm we present. The choice of reasonable ranges from which the source parameters are drawn is also discussed.

%************************************************************
\subsection{\label{Sec:DataPro} Data Production}
%***********************************************************

We have generated one month of gapeless gaussian detector noise data assuming the expected best strain sensitivity of advanced LIGO-Virgo~\cite{ADvLIGO,AdvVirgoRef:2009} detectors, i.e. a noise spectral density $S_h = 4 \times 10^{-24}$~Hz$^{-1/2}$ in the frequency interval $[70,\, 200]$~Hz. By using the argument detailed in Sec.~\ref{FFTfracOrbit}, we have produced 10\,127 interlaced FFTs with duration $\Tfft = 512$~s each. Then, we artificially generated and added to such a data set 131 CW signals emitted from pulsars in low- and high-eccentricity binary systems.
Software-injections have been performed using the LALSuite software package~\cite{LALSuite}.
As already anticipated, we have neglected spin wandering effects as the current method -due to its robustness- has a high tolerance to small frequency variations, and then is expected not to be limited by possible spin wandering.

%************************************************************************
\subsection{\label{Sec:SignParam} Choice of signal parameters}
%*************************************************************************

\begin{figure}[htbp]
  \centering
  \includegraphics[width=\columnwidth]{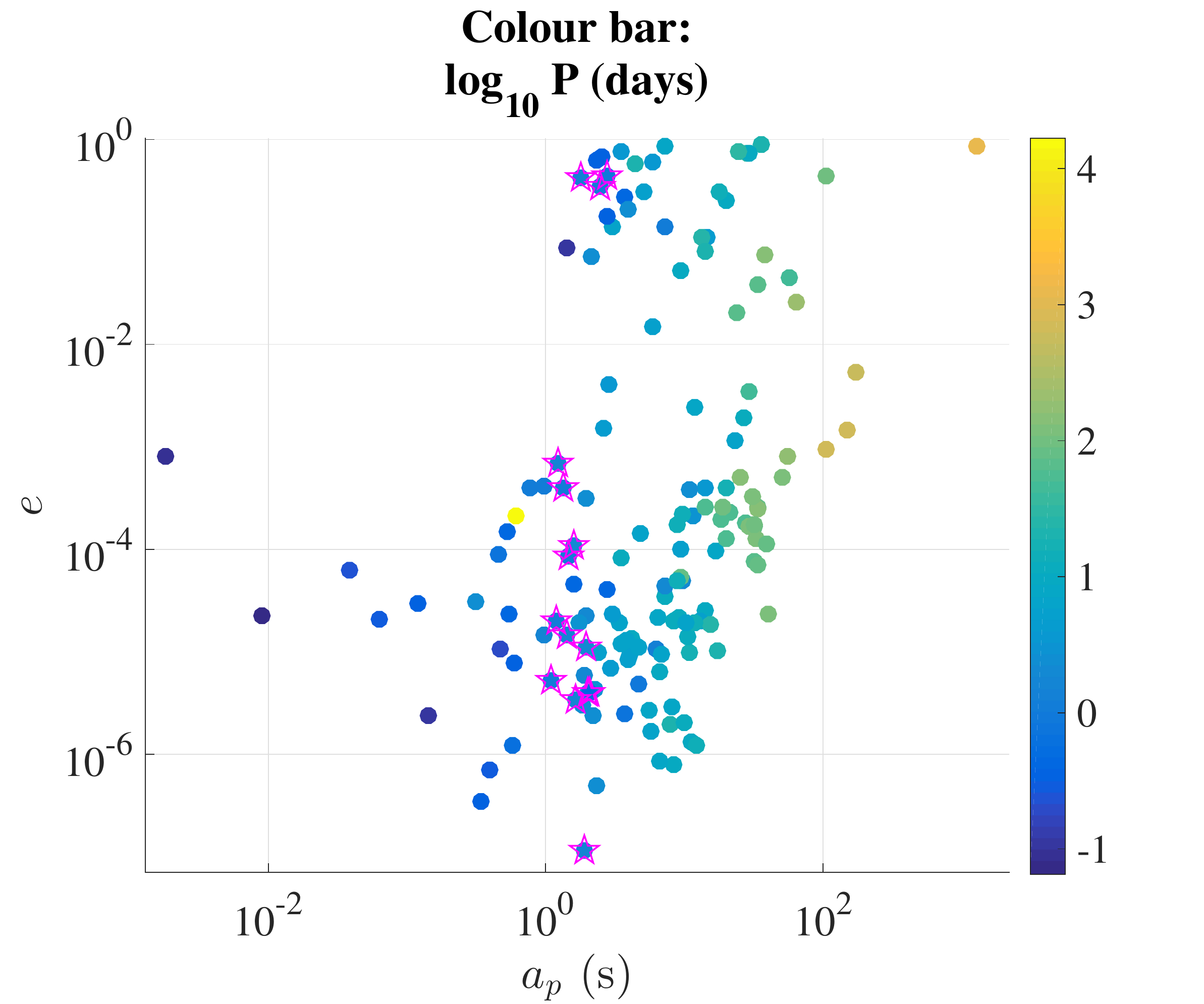} 
  \caption{(Color online) Distribution of orbital eccentricities versus projected semi-major axis for 221 ATNF catalogue pulsars
found in binary systems with rotation frequencies $\geq$10 Hz, and up to $\sim 716$~Hz~\cite{atnf,Manchester:2004bp}. The colour bar indicates 
the logarithmic values of source orbital periods. The points surrounded by magenta stars correspond to sources with $10~\mathrm{h} \leq P\leq2$~d, $1~\mathrm{s} \leq  \asini\leq3$~s, and $0 <  \ecc < 1$.}     \label{fig:apEccBinPulsATNF}
\end{figure}

In Fig.~\ref{fig:apEccBinPulsATNF} we plot the distribution of orbital eccentricities versus projected semi-major axis for 221  ATNF catalogue pulsars
found in binary systems. The minimum and maximum orbital periods in Fig.~\ref{fig:apEccBinPulsATNF} are roughly 1.6~h and 46~y, respectively. 
The binary systems we are interested in, however, have orbital periods $10~\mathrm{h} \leq P\leq2$~d, projected semi-major axis $\asini$ of a few seconds, and no restriction on orbital eccentricity (see magenta stars in Fig.~\ref{fig:apEccBinPulsATNF})  \footnote{We note that such class of signals includes of course Scorpius~X-1.}.

The so-called signal \textit{amplitude parameters} are randomly chosen from uniform distributions as follows:
the scalar gravitational-wave amplitude $h_0^s \in [1, 5]\times 10^{-24}$, the inclination angle $\cosisour \in[-1,1]$, the polarization angle within
$\psi_s\in[0,2\pi]$, and the (irrelevant) initial phase within $\phi_{0_s} \in[0,2\pi]$\footnote{The initial phase $\phi_0$ at the reference time $\tref$ has to be added in Eqs.~\eqref{eq:ComplLinPhaseApprox} and~\eqref{eq:EccAnPhase}, but for our purposes it can also be neglected.}.

The sky-position for all signals is fixed to that of Scorpius~X-1, namely $(\alpha_s,\,\delta_s) = (4.276,-0.273)\,$rad~\footnote{In order to avoid sensitivity losses from sky-positions which are less favorable at certain times, we could normalize the selected peaks with \textit{ad-hoc} weights based on the so-called antenna pattern, i.e. the directional (angular) response of the detectors. This could be done in a way analogous to what is implemented in~\cite{Astone:2014esa}.}.

The so-called signal \textit{phase-evolution parameters} $ \{f_s, \asinisour,\,\tPerisour,\,P_s,\, e_s,\,\argpsour \}$ are generated by randomly drawing them from uniform distributions over the ranges:
\begin{align}
f_s & \in [70, 200]\, \mathrm{Hz} \notag\\ 
\asinisour &\equiv \frac{a\,\sin{}i}{c} \in [1,\,3]\,\mathrm{s}, \notag\\
 P_s &\in [10,\, 48]\,\mathrm{h},\notag\\
 \tPerisour &\in \left[\tMid-\frac{P}{2},\,\tMid + \frac{P}{2}\right]\,,\label{eq:RandSpurParam}\\
 \log_{10} e_s &\in [-6,\,\log_{10}(0.9)]\,,\notag\\
 \argpsour &\in [0,\,2\pi]\,\mathrm{rad}\,,\notag
\end{align}
with $\tMid$ being the midtime of the whole observation.

We remark that we consider also high-eccentricity orbits, favouring low-eccentricity ones as more copious (see Fig.~\ref{fig:apEccBinPulsATNF}).

Every frequency $f_s$ has been actually uniformly drawn from a subinterval of 0.4~Hz around the mid interval of every analysed 1~Hz band (whose choice is motivated a few lines later). This is done to avoid border effects\footnote{In presence of signals whose frequency modulation spans more than one frequency bin, the signal SNR would be lessened in every bin.}, which would make  borderline detections more difficult. A way to circumvent this issue consists of performing a trivial interlacement of all analysed frequency intervals, which will be anyway done in a real search. This would increase the computing cost, as it would double the number of analysed frequency bands, but will avoid to underestimate signal amplitudes and miss borderline detections. 
Due to the paltry computing time, such an increase will anyway keep the method quite computationally feasible even on a single-processor computer (see Sec.~\ref{Sec:TeoCtot}). 

\begin{figure}%[h!]
  \centering
  \includegraphics[width=\columnwidth]{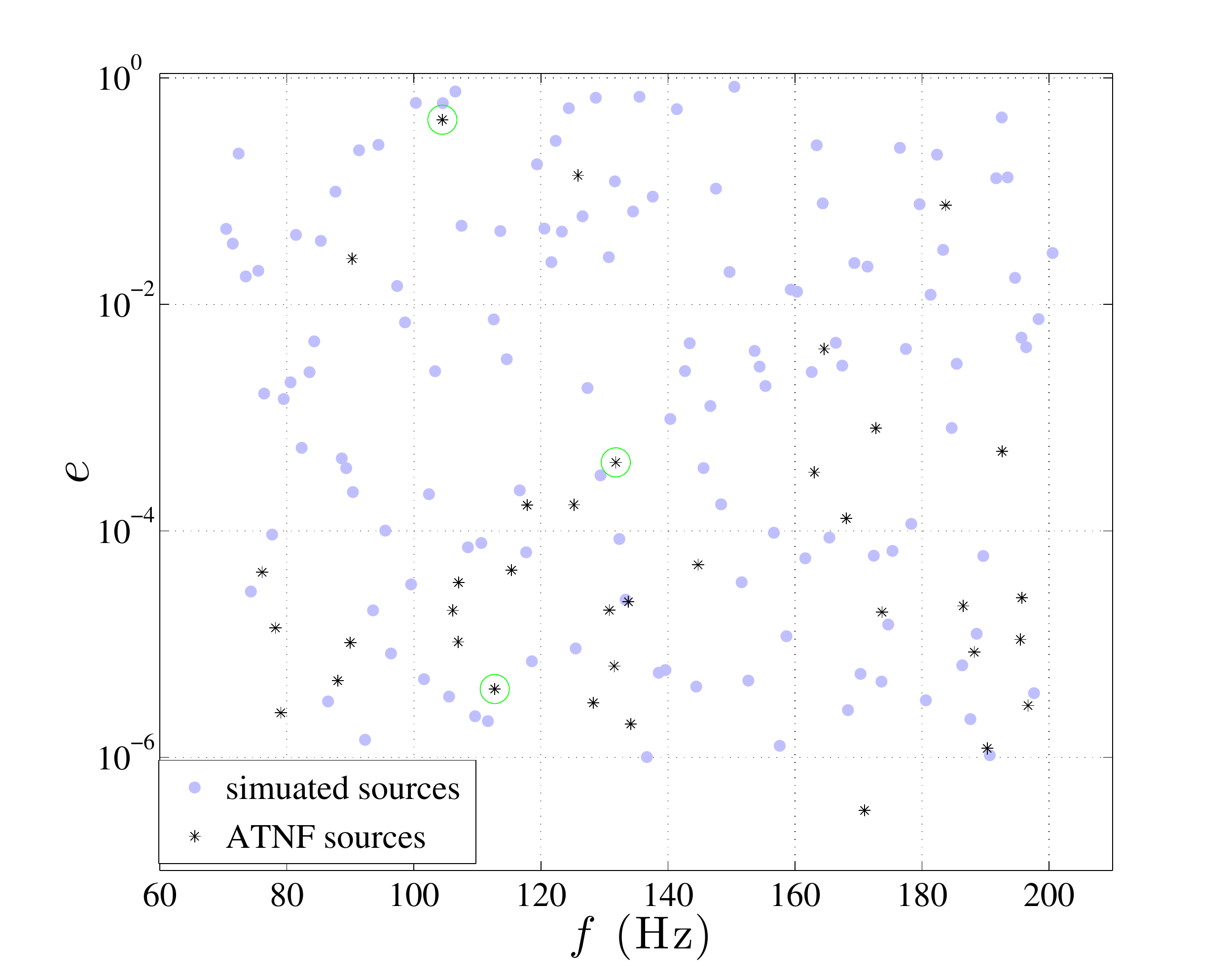} 
  \caption{(Color online) Distribution of orbital eccentricities versus frequency for the population of 131 simulated sources (dots) and 45 ATNF catalogue pulsars
found in binary systems (asterisks). The asterisks surrounded by empty circles correspond to sources with orbital periods and projected semi-major axes compatible with our target population.}
    \label{fig:EccMieATNF}
\end{figure}

In Fig.~\ref{fig:EccMieATNF} we show the orbital eccentricities versus frequency for 131 simulated sources (light dots) and 45 known pulsars found in binary systems (dark asterisks)~\cite{atnf,Manchester:2004bp} in the same frequency and eccentricity ranges of the simulated population. The asterisks surrounded by circles indicate sources with $10~\mathrm{h} \leq P\leq2$~d and $1 \leq \asini \leq3$~s, which is our target population. Although they are only three sources, there are several other pulsars we expect to exist in the Galaxy, which have not yet been discovered (as discussed in Sec.~\ref{Intro}). Hence, it becomes crucial to sift a parameter space as much large as possible, without exclusion of particular regions.

As stated later, we split the searched frequency range into smaller frequency bands that will be deeply post-processed to find evidence of CW signals. We choose 1~Hz as width to analyse the various frequency bands, which is much larger than the maximum modulation caused by the orbital motion of the source in the parameter space investigated here. In fact, Fig.~\ref{fig:MaxModDepth} exhibits the maximal Doppler shift due to orbital motion (discussed in Appendix~\ref{sec:maxim-doppl-shift}), i.e. $2\, f_s\,\Delta M$ [see Eq.~\eqref{eq:modDepth}], versus the frequency of 131 simulated sources. We see that $2\, f_s\,\Delta M$ varies from $8.2\times 10^{-3}$~Hz up to $\sim 0.2$~Hz. 
\begin{figure}[htbp]
  \centering
  \includegraphics[width=\columnwidth]{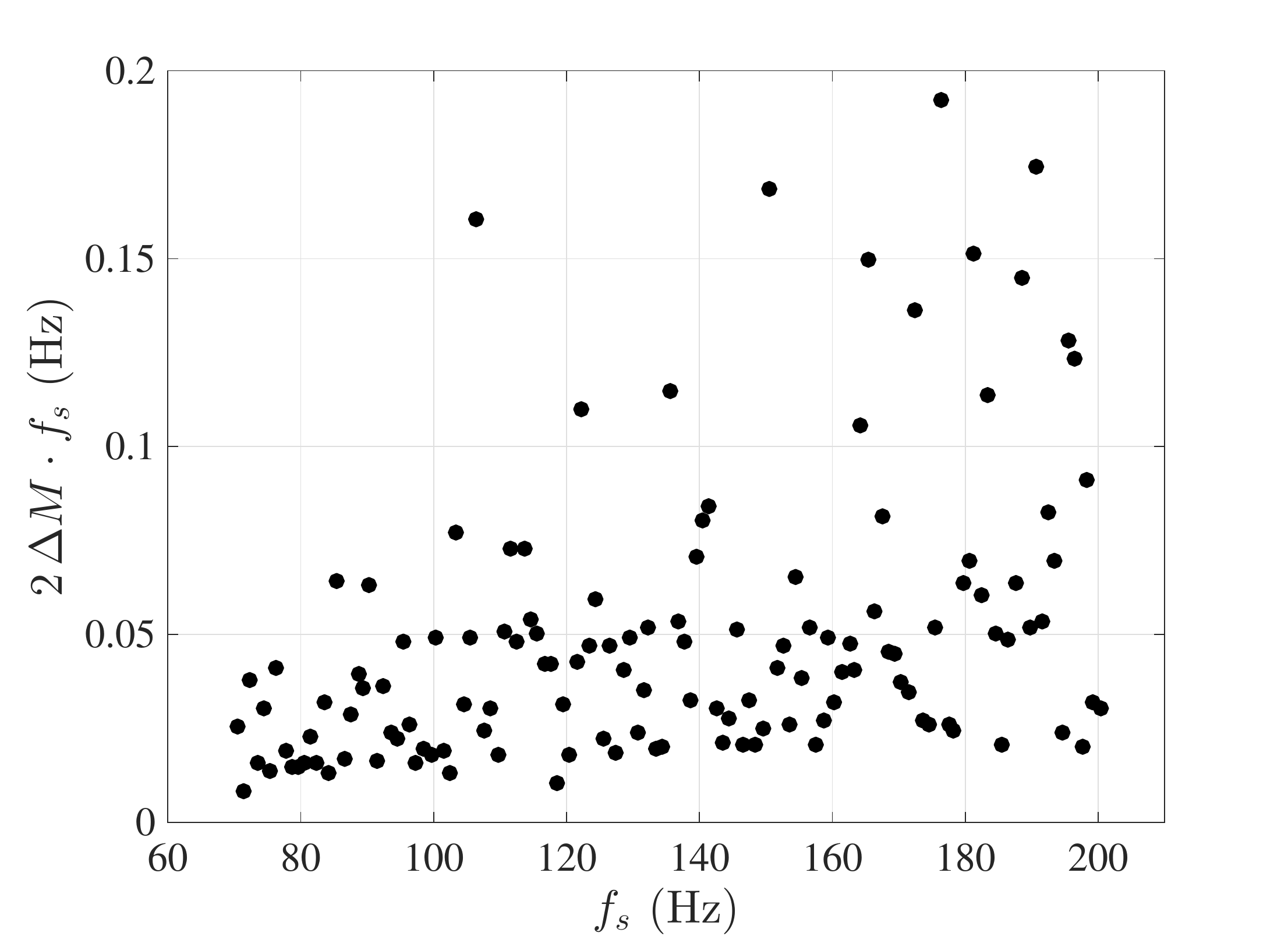}
  \caption{Maximal Doppler shift due to orbital motion ($2\,f_s\,\Delta M$) as a function of the frequency for the population of 131 simulated sources.} 
    \label{fig:MaxModDepth}
\end{figure}
%****************************************************************************************************
\section{\label{FFTfracOrbit} Constraining the FFT duration}
%****************************************************************************************************

The time baseline of an FFT is typically determined by requiring the signal power to be concentrated in less than a frequency bin. In Appendix~\ref{Sec:Tsft-choice} we have nailed down the longest possible FFT duration based on the phase error of the linear-phase approximation over an FFT. The estimate provided in Eq.~\eqref{eq:FFTlen} is, however, very conservative. Indeed, assuming a phase error $\Delta \phi = \pi/4$, and the most unfavourable boundaries of the investigated parameter space, i.e. $P=10$~h, $e=0.9$, $a_p=3$~s, $f=200$~Hz, we obtain $T_\mathrm{FFT}\sim129$~s. 
The FFT durations we used to process the set of data containing 131 sources, simulated with parameters within the ranges provided in Eqs.~\eqref{eq:RandSpurParam}, are instead $512$~s, which are based on an approximation that allows us to improve the sensitivity by a factor of $\sqrt{512/129}\sim 2$.

The reasoning behind the usage of FFTs longer than those obtained with Eq.~\eqref{eq:FFTlen} consists of considering, rather than the whole orbit swept by a source, only a large fraction of it, i.e. at most the 80\%. 
This is somewhat reasonable as, over one month, we will observe from a minimum of 15 orbits to a maximum of 72 orbits (considering $10~\mathrm{h} \leq P\leq2$~d).
We explain better how we achieved such a consideration.
We performed simulations for $1\,000$ sources, with random parameters drawn uniformly from the ranges of Eqs.~\eqref{eq:RandSpurParam}, and for each source we compute the rate at which the source changes its velocity by taking the second time derivative of Eq.~\eqref{RoemD} (i.e., the acceleration divided by the light speed). We then choose, for every source, the 80th percentile of $|\ddot{R}(t)|/c$, neglecting thus the 20\% highest values in terms of $|\ddot{R}(t)|/c$, which correspond to parts of the orbit where there is the largest source velocity variation, and then the largest orbital Doppler modulation. 

\begin{figure}[htbp] 
\centering
\includegraphics[width=1.2\columnwidth,clip]{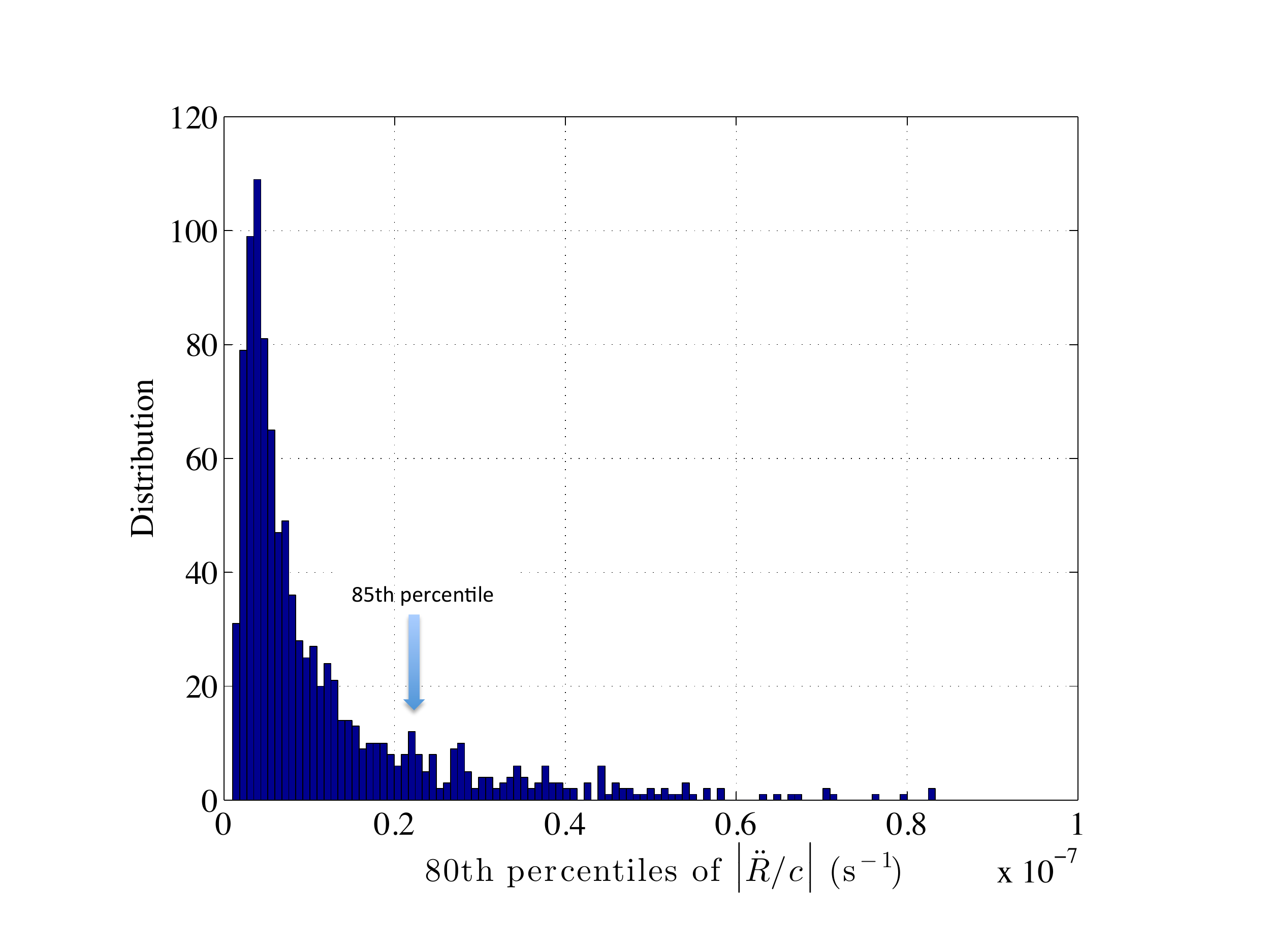}
\caption{(Color online) Distribution of the 80th percentiles of $|\ddot{R}(t)|/c$ for 1\,000 sources. The arrow denotes the value of $|\ddot{R}(t)|/c$ corresponding to the 85th percentile, i.e. $|\ddot{R}_0.85|/c\sim 2 \times 10^{-8}$~s$^{-1}$.\label{Fig:HistOf80thPercentiles}}
\end{figure}

Figure~\ref{Fig:HistOf80thPercentiles} shows the distribution of the 80th percentiles of $|\ddot{R}(t)|/c$ for 1\,000 simulated sources.
We see that the maximum value of the 80th percentiles of $|\ddot{R}(t)|/c$ is $\mathcal{P} \sim 8 \times 10^{-8}$~s$^{-1}$, which is of course the 100th percentile of the distribution in Fig.~\ref{Fig:HistOf80thPercentiles}. We should use such value to reliably estimate the maximum FFT duration as a function of the search frequency:
\begin{equation}
  \label{eq:longerTfft}
\Tfft^{\mathrm{max}} =  \frac{1}{\sqrt{\mathcal{P}\, f}}. 
\end{equation}
If we consider instead the 85th percentile (marked by the arrow in Fig.~\ref{Fig:HistOf80thPercentiles},) i.e.  $ \mathcal{P}_{0.85}  \sim 2 \times 10^{-8}$~s$^{-1}$, we can further lengthen the FFT duration by a factor 2 at the expense of further reducing the fraction of the orbit for 15\% of the sources, considering for them even smaller velocity variations. 

Hence, replacing $\mathcal{P}$ with $\mathcal{P}_{0.85}$ in Eq.~\eqref{eq:longerTfft}, we obtain $\Tfft^{\mathrm{max}}=512$~s, which we use to produce FFTs for the whole search frequency range $f \in [70, 200]\, \mathrm{Hz}$.
Such a choice is actually based on the (worst) highest search frequency, i.e.,~200~Hz, as for~70~Hz an FFT of $845$~s would be more appropriate. As this translates into a reduction in sensitivity (and then in parameter estimate accuracy), one can envisage to produce SFDB of different duration, depending on the highest frequency value of the subinterval into which the entire search frequency range has been previously split.

%****************************************************************************************************
\section{\label{Sec:method} The search method } 
%****************************************************************************************************
The new strategy presented here relies on the basic consideration that the frequency modulation caused by the source orbital motion can be used to unveil the signature of CW signals, and to also extract information on source parameters. We summarise the salient steps of the procedure in the following.

\begin{enumerate}
\item The starting point of the analysis are the subpeakmaps obtained by processing the artificial data set generated according to the guidelines described in Sec.~\ref{Sec:PS}. In particular, we have $\mathcal{N}=10\,127$ subpeakmaps and interlaced FFTs of duration $\Tfft=512$~s, covering an overall one month of gapeless data.

\item For every subpeakmap, for the source sky location assumed to be known with enough accuracy, we apply a frequency correction to account for the Doppler shift due to the Earth motions. Hence, neglecting relativistic effects, we shift the peak frequencies $f_p$ from the original received ones, $f_p^o$, according to
\begin{equation}
f_p = \frac{f_p^o}{\left(1 + \frac{ \vec{v} \cdot \vec{n} }{c}\right)},
\end{equation}
where $\vec{n}$ is the unit vector pointing from the SSB to the source (see Sec.~\ref{Sec:signal});
 $\vec{v}$ is the detector velocity with respect to the SSB frame, and is given by the sum of two components, from the yearly Earth motion around the Sun and from the rotation of Earth around its axis. These velocities are computed at the FFT midtimes ($t_c$) by using the DE405 JPL Solar System Ephemeris~\cite{Standish82,Standish90}.

We emphasize that $f_p$ are then the peak frequencies modulated due to the only source orbital motion, which is the information we need for the current procedure. 

\item As previously stated, we analyse the search frequency range splitting it into 1~Hz frequency bands, and each of them undergoes scrutiny to establish if the modulated pattern of a CW signal is present. To this purpose we apply the filters described in Sec.~\ref{Sec:GaussFilter} and, if a signal is found, we extract the subband containing the modulated pattern that needs  additional inspection.

\item In every subband identified as above, we select the most significant peaks crossing the threshold established in Sec.~\ref{Sec:ThresholdChoicePA} (i.e. having $\mathcal{R} > \theta_\mathrm{thr}$).

\item We then average the frequencies of the most significant peaks identified at the previous stage, and corresponding to a same FFT midtime, resulting at most in $\mathcal{N}$ $(t_c, \bar{f}_p)$ pairs per subband~\footnote{In the current analysis we found a minimum (maximum) of 14 (9\,417) pairs $(t_c, \bar{f}_p)$, in the band [151,\,152]~Hz ([82,\,83]~Hz), where a signal with an amplitude $h_0 = 1.1 \times 10^{-24}$ ($h_0 = 4.3\times 10^{-24}$) has been injected. Special care must be devoted not to consider such a number of FFTs as those contributing to detect a putative signal, but rather to estimate the parameters of a previously identified signal. In fact, a detection claiming depends on the performance of the filters introduced in Sec.~\ref{Sec:GaussFilter} (as clearly stated later), and applied before selecting peaks with $\mathcal{R} > \theta_\mathrm{thr}$.}.

\item For each subband, we perform a periodogram estimate (detailed in Sec.~\ref{Sec:PeriodogramEstim}) for the unevenly spaced data set $(t_c, \bar{f}_p)$ in order to look for periodicities, and possibly estimate source orbital period.

\item We then carry on with a least-squares fitting of sine waves to estimate the signal frequency and remaining orbital parameters (as depicted in Sec.~\ref{Sec:LSF}).
\end{enumerate}

Further details follow.

%****************************************************************************************************
\subsection{\label{Sec:GaussFilter} Identifying Signal Pattern}
%****************************************************************************************************

To identify the pattern of a continuous wave signal, whose frequency is modulated by the source orbital motion,
the peak frequencies in every 1~Hz analysed frequency band undergo a set of four filters in cascade, described in the following.

\begin{figure*}[htbp] 
  \raggedright (a)\hspace*{\columnwidth}(b)\\[-0.03cm]
  \includegraphics[clip,width=\columnwidth]{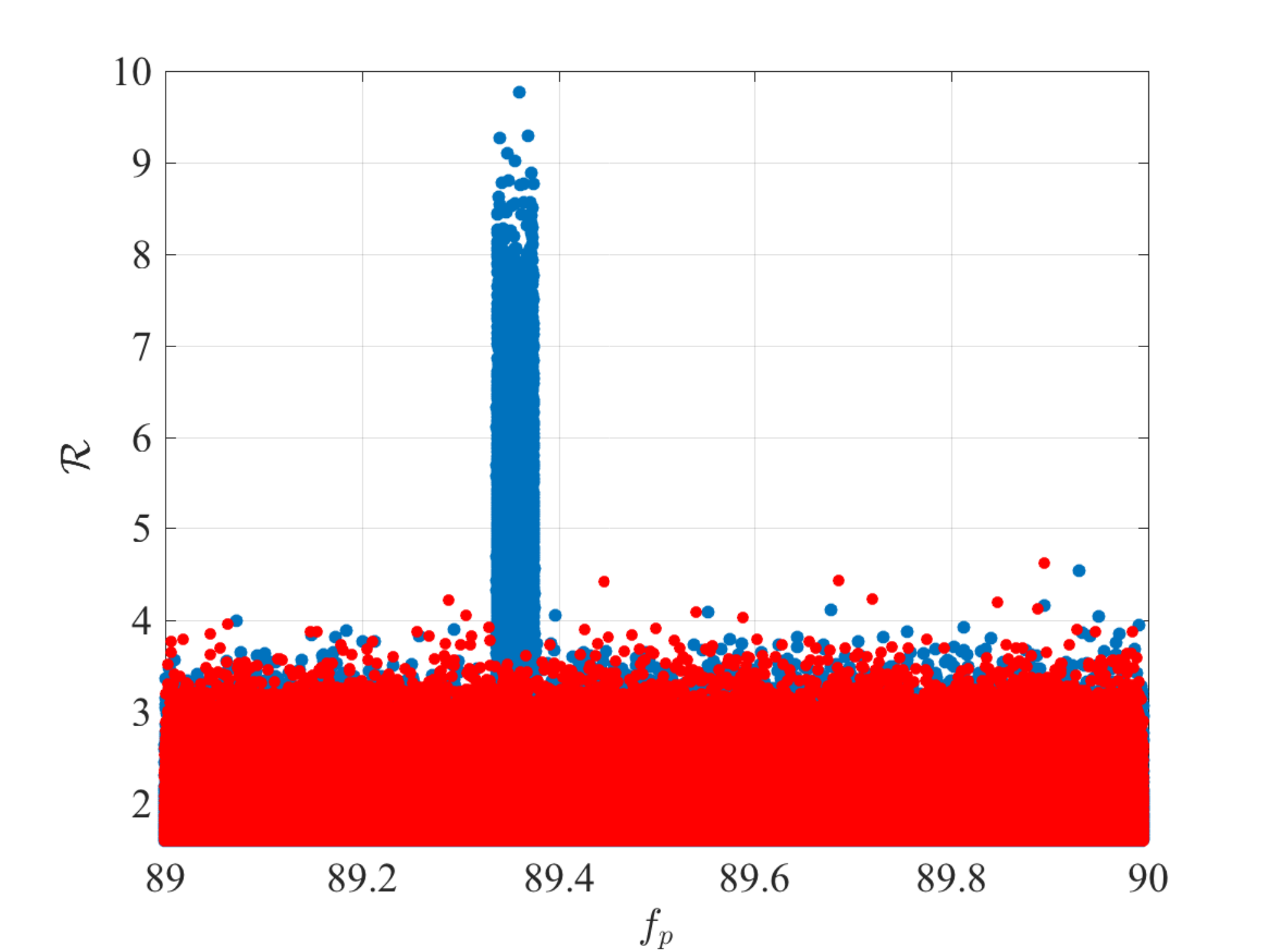} 
  \includegraphics[clip,width=\columnwidth]{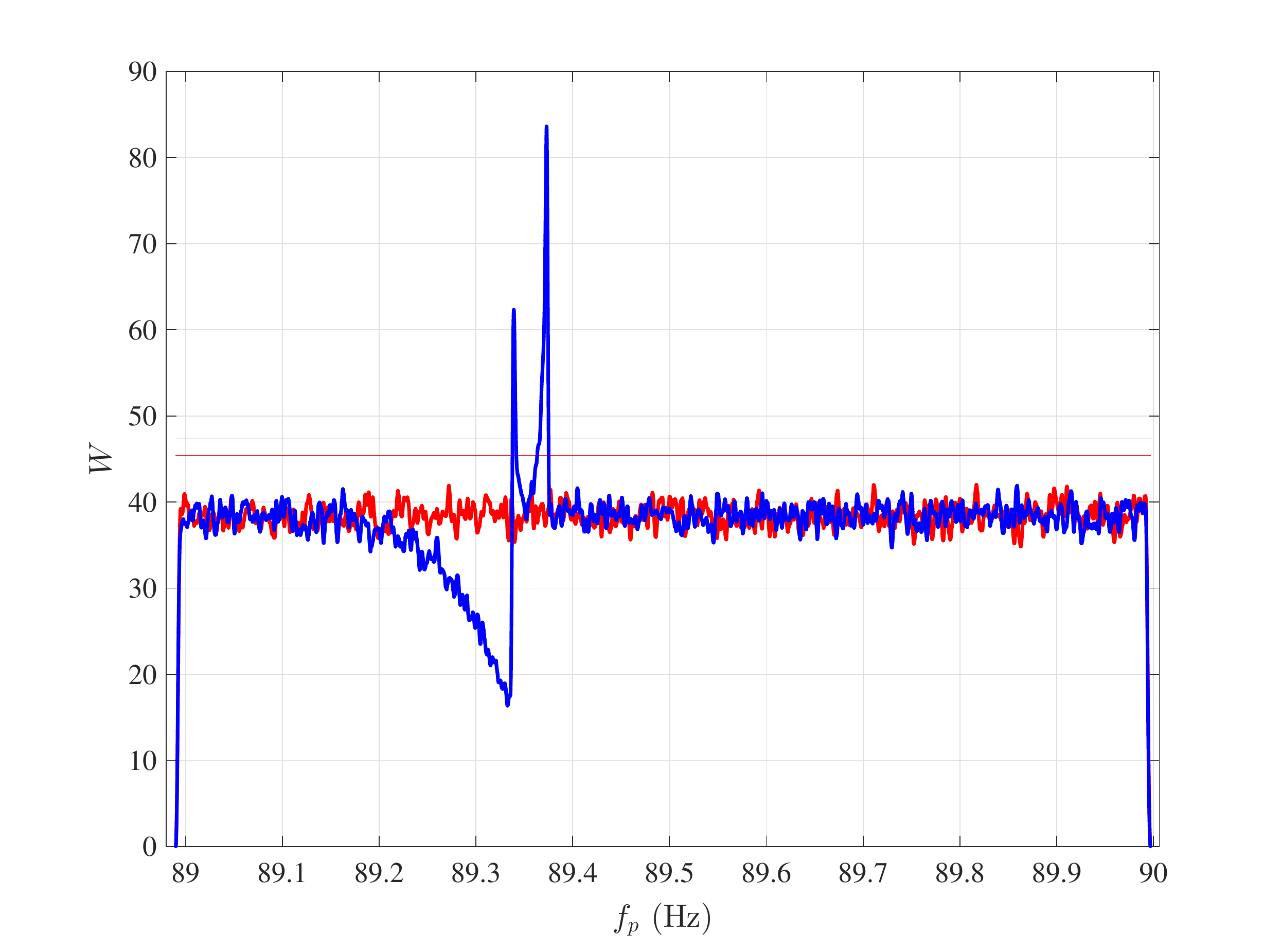}\\[-0.03cm]

  \raggedright (c)\hspace*{\columnwidth}(d)\\[-0.03cm]
  \includegraphics[clip,width=\columnwidth]{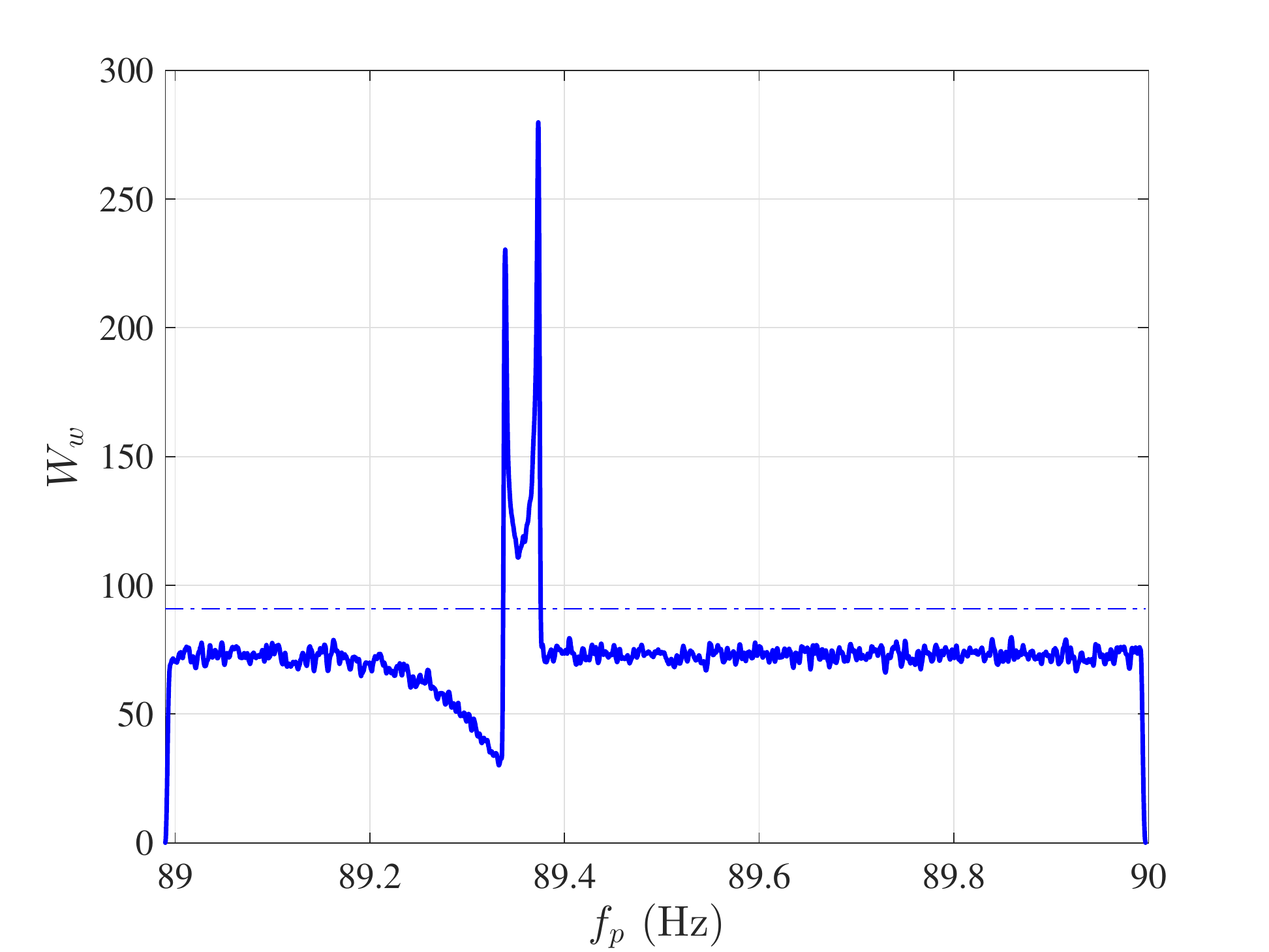}
  \includegraphics[clip,width=\columnwidth]{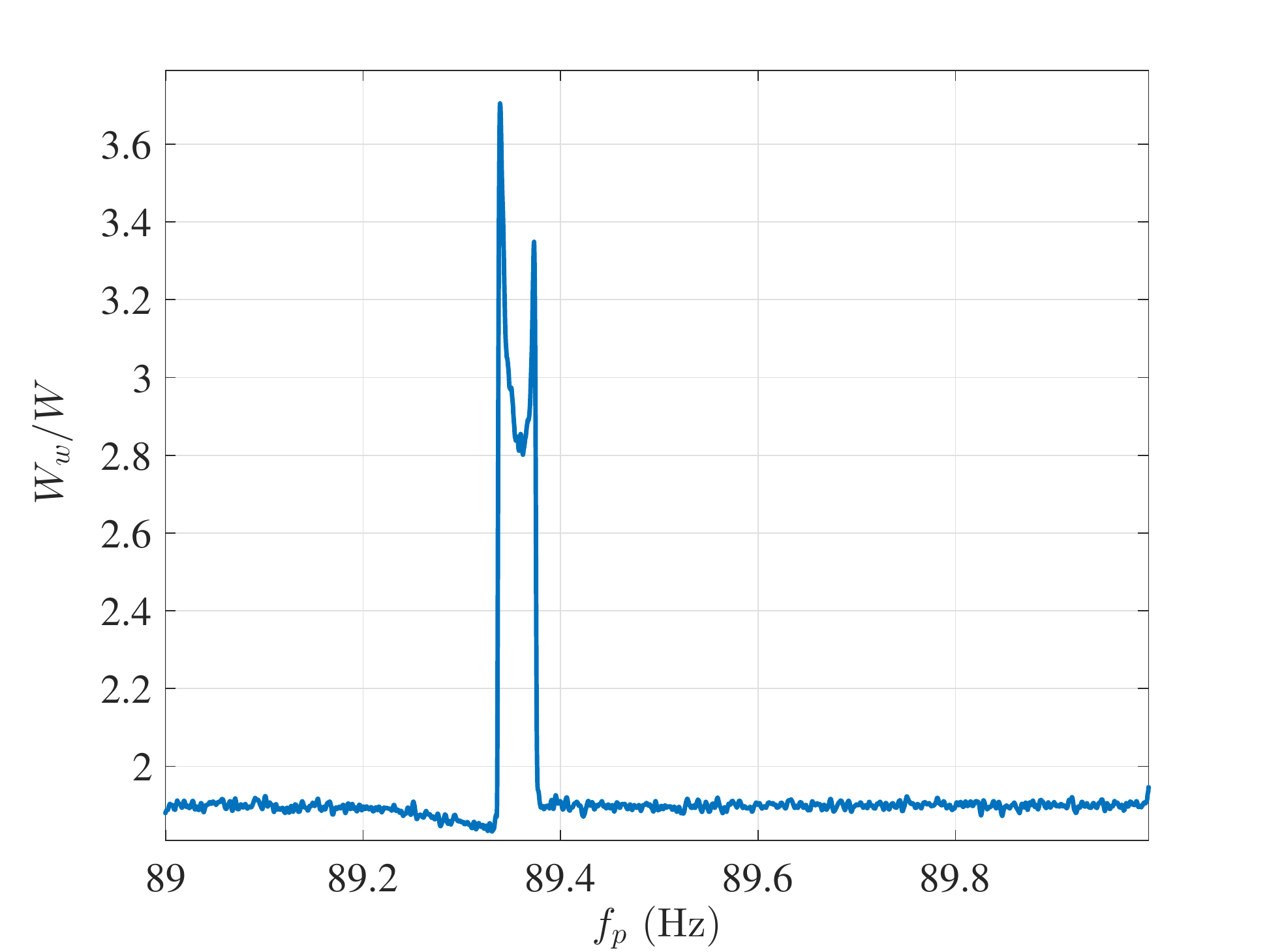}\\[-0.3cm]

  \caption{(Color online) Panel (a): Peak amplitude $\mathcal{R}$ versus peak frequencies for two frequency bands where a modulated CW signal is present (blue dots), with an orbital Doppler modulation $2\,\Delta M\,f_s \sim 37$~mHz , and absent (red dots). Panel (b): Output $W$ of a triangular impulse response filter applied to the peak frequencies in a band where there is a signal added into Gaussian noise data (blue curve), and the same band when no signal is present, but just pure Gaussian noise (red curve). The highest and lowest lines correspond to the value $m(1) + 6 \, m(2)$ in the two cases, respectively. Panel (c): Output $W_w$ of the triangular impulse response filter applied to the peak frequencies weighted by their respective peak amplitude; the dashed line corresponds to the value $m_w(1) + 6 \, m_w(2)$ (see main text for the definition of the $m$ and $m_w$ values.). Panel (d): Ratio of $W_w/W$ versus peak frequencies.}
 \label{fig:TriangularFilterResp}
\end{figure*}

Figure~\ref{fig:TriangularFilterResp}~(a) shows the peak amplitude $\mathcal{R}$ as a function of the peak frequencies for two superimposed frequency bands taken as examples where a modulated CW signal is present (blue dots) and absent (red dots). In order to facilitate the task of identifying a signal, we first filter the peak frequencies by using a standard triangular impulse response filter (i) [whose output is denoted with $W$ in Fig.~\ref{fig:TriangularFilterResp}~(b)], and a window half-width equal to $\Tfft^{-1}$ (empirically chosen). As clarified later, we apply the same filter to the peak frequencies weighted by the peak amplitude $\mathcal{R}$ (ii), obtaining an output $W_w$ [shown in Fig.~\ref{fig:TriangularFilterResp}~(c)]. These filters are implemented in the SNAG software package~\cite{SNAG}, with the possibility of choosing either a triangular, rectangular or exponential window, which exhibit similar results when applied to the same data set. In the following studies we opted however for the most intuitive choice of a triangular impulse response. 

In order to first identify frequency bands with potential signals standing out from noise, we verify if

\begin{equation}
  \label{eq:NfromSt}
    \left\{\begin{array}{ll}   \max{W} > m(1) + 6 \, m(2)\\
    \max{W_w} > m_w(1) + 6 \, m_w(2),\\
  \end{array} \right.
\end{equation}
where
\begin{equation}
  \label{eq:RobustMedian}
 m(2) = \frac{\mathrm{median}( |W - m(1)| )}{\mathcal{C}},
\end{equation} 
with $m(1)=\mathrm{median}(W)$, and $\mathcal{C}=0.6745$ is a normalisation factor such that, if the distribution of $W$ is normal, then $m(2)$ is the standard deviation~\cite{Astone:2014esa}. 
We note that $m_w(1)$ and $m_w(2)$ have the same meaning of $m(1)$ and $m(2)$, respectively, but are referred to $W_w$.
We use the median, rather than the mean, as it is more robust in the presence of outliers, and we choose it to build a robust estimator of the dispersion parameter, which we use instead of the classical standard deviation. 

Figure~\ref{fig:TriangularFilterResp}~(b) shows the output of the triangular impulse response filter (i) applied to a 1~Hz-wide frequency band where a modulated CW signal is present (blue curve), and absent (red curve). In this last case of pure Gaussian noise there are no outliers crossing the red dashed line, which correspond to the value  $m(1) + 6 \, m(2)$. 

We stress that the criterion given in Eq.~\eqref{eq:NfromSt} is not sufficient to prevent the selection of instrumental artefacts, but it is used to quickly sift the analysed frequency bands, and identify the disturbed ones, which will be inspected with more scrutiny. Furthermore, in the present work we have simulated a single detector data set adding fake signals into Gaussian noise data, but when applying the current method to real case, data taken from multiple detectors will be available, allowing us to more reliably exclude prominent disturbances by using a coincidence analysis. The application of more aggressive noise identification and artefact mitigation techniques is also envisaged to be beneficial, and is currently under investigation.

For all frequency bands satisfying Eq.~\eqref{eq:NfromSt}, we compute the ratio of $W_w/W$, illustrated in Fig.~\ref{fig:TriangularFilterResp}~(d). This is done to remove the depletion visible in the blue curves of Fig.~\ref{fig:TriangularFilterResp}~(b)~and~(c), which is due to the peak selection effect in the peakmap~\cite{Astone:2005fj,Astone:2014esa}, and is more prominent for loud signals. To enhance the signal contribution, the ratio $W_w/W$ is convolved against a filter (iii) with a non-symmetrical Gaussian shape response, and finally a filter identical (iv), but running in opposite direction. This is needed to reproduce, in the plane ($f,\,\mathcal{R}$), the shape of a signal whose frequency is modulated by the source orbital motion [see blue dots in Fig.~\ref{fig:TriangularFilterResp}~(a)].

\begin{figure*}[htbp] 
  \raggedright (a)\hspace*{\columnwidth}(b)\\[-0.03cm]
  \includegraphics[clip,width=\columnwidth]{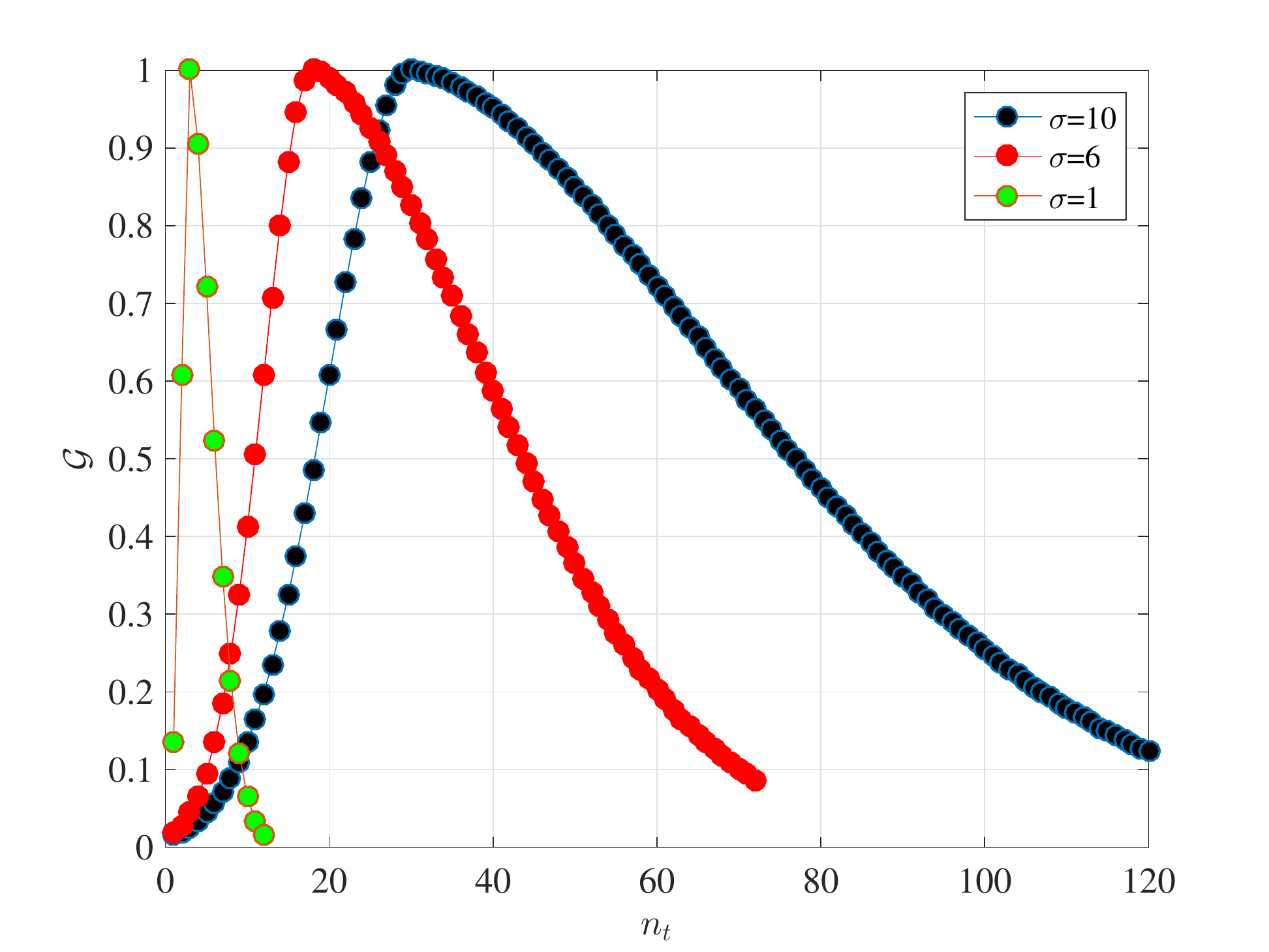} 
  \includegraphics[clip,width=\columnwidth]{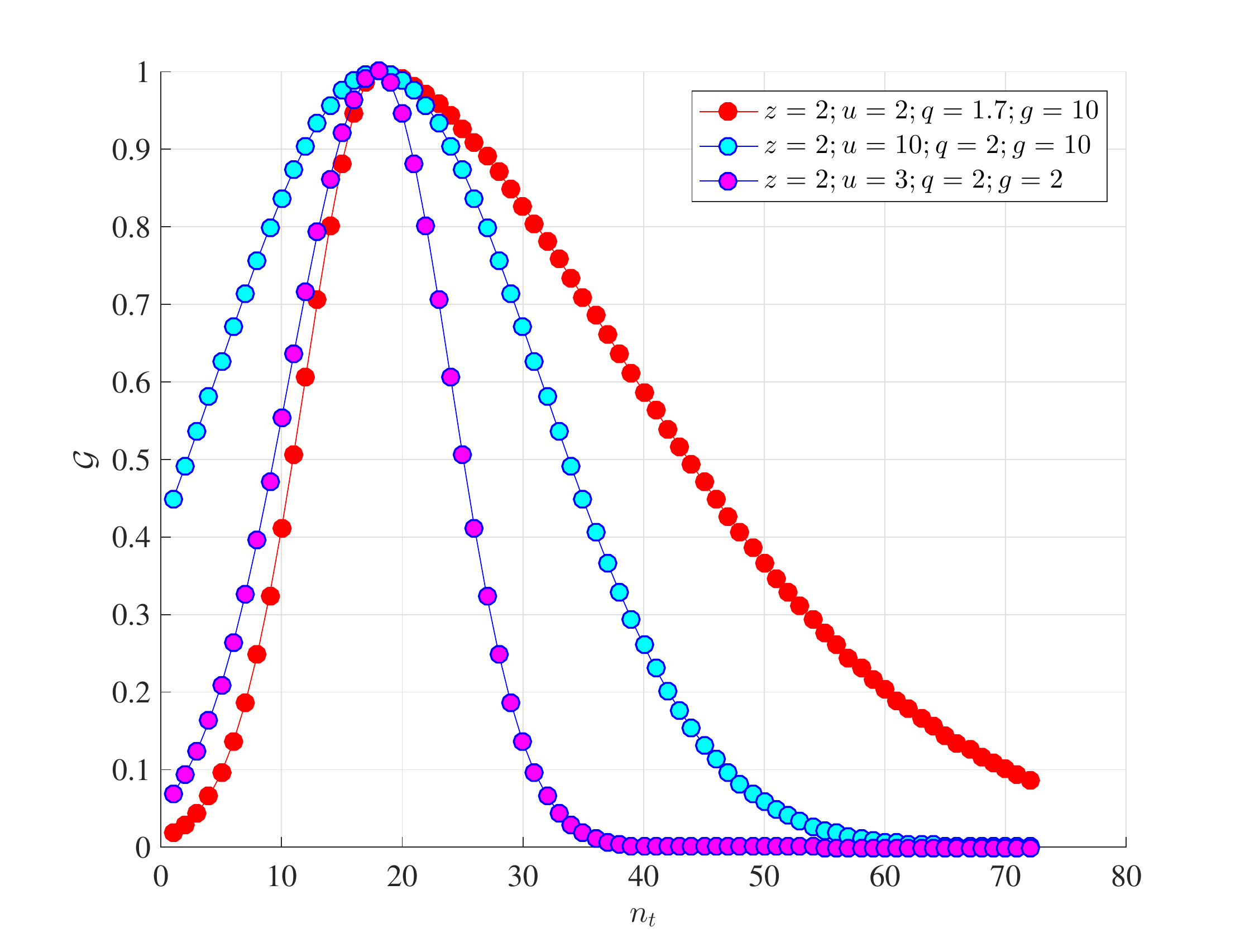}\\[-0.03cm]
  \caption{(Color online) $\mathcal{G}$ function of Eq.~\eqref{GaussFilter} versus $n_t$ for $z=2,\,u=2,\,q=1.7\,,g=10$ and three $\sigma$ values [panel~(a)], and $\sigma=6$ with different values of $z,\,u,\,q\,,g$, shown in the legend [panel~(b)].}
 \label{fig:GNfilter}
\end{figure*}

Such a Gaussian like filter, modelling half part of the horn-shaped signal modulation pattern, is given by
\begin{equation}\label{GaussFilter}
  \left\{\begin{array}{ll} \mathcal{G}(1:\mu) = e^{-\frac{(t(1:\mu) - \mu)^z}{u\,\sigma^2}}\\
 \mathcal{G}(\mu+1:n_t) = e^{-\frac{(t(\mu+1:n_t)-\mu)^{q}}{g\,\sigma^2}},\\
\end{array} \right.
\end{equation}
 where $\mu = 3\,\sigma$, $n_t=12\,\sigma$, $z=2$, $u=2$, $q=1.7$, and $g=10$ (all values found empirically). The $\mathcal{G}$ function is plotted in Fig.~\ref{fig:GNfilter}~(a) for different $\sigma$ (i.e. ``standard deviation") values and $z=2,\,u=2,\,q=1.7\,,g=10$, while in Fig.~\ref{fig:GNfilter}~(b) for different values of $z,\,u,\,q\,,g$ and $\sigma=6$. 
 
 Simulation-based studies bring us to choose a $\mathcal{G}$ function with $z=2,\,u=2,\,q=1.7\,,g=10$, and $\sigma=6$ (red curve in Fig.~\ref{fig:GNfilter}).
 
To model the remaining half part, the convolution is computed by using the same Gaussian like filter running in opposite direction. 
In general, if a signal is present in a band, when the Gaussian like filter runs in one direction, only one of the horns in Fig.~\ref{fig:TriangularFilterResp} gets amplified, and the other horn will be amplified by the filter running in opposite direction. If this condition is not satisfied, we consider the signal \textit{not found} in the analysed frequency band. On the contrary, we can \textit{claim a detection} and identify the band around the modulated signal pattern. For safety reasons we select, however, a wider band consisting of roughly 100 additional samples (corresponding to $\sim10$~mHz) on both sides.

The standard deviation of the Gaussian like filter $\sigma=6$ is chosen to be wide enough to account for the Doppler shift due to source orbital motion. If a signal is not found in a frequency band, however, a second attempt is done by convolving the data against an identical Gaussian like filter, but with $\sigma=1$, to take into account also smaller Doppler modulation effects. 

After selecting the frequency band around the modulated signal pattern, we proceed on setting a threshold according to what explained in Sec.~\ref{Sec:ThresholdChoicePA} to select only the most significant peaks within such a subband.

The false alarm probability for the identification of a signal (i.e., a false alarm) in frequency bands where no signal is present is computed generating 60\,000 pure Gaussian noise realisations for which we verify if the condition expressed by Eq.~\eqref{eq:NfromSt} is satisfied. Since this is not fulfilled, we can place an upper limit on the false alarm probability, which results being smaller than $1.6 \times 10^{-5}$. We note that such value is very conservative as it does not take into account that a false alarm, if found, will undergo further checks via the Gaussian like filter before being claimed a detection (according to what explained above).

We note that the signal visible in Fig.~\ref{fig:TriangularFilterResp}~(a) has a strain amplitude of $\sim 2.7 \times 10^{-24}$ over a $\Tobs = 30$~days. As shown in Sec.~\ref{Sec:SensEstim}, we would have detected the same signal also if it would have had a strain amplitude of $\sim 8.7 \times 10^{-25}$ over the same observation time.

\newpage
%****************************************************************************************************
\subsection{\label{Sec:ThresholdChoicePA} Setting stringent threshold for peak selection}
%****************************************************************************************************

The threshold $\theta_{\mathrm{thr}}$ we choose to select the most significant peaks affects the parameter estimate abilities and search sensitivity. The criterion we use for the choice of  $\theta_{\mathrm{thr}}$ is the maximisation of an informative observable $\Phi(\theta)$, which combines the number of peaks solely due to signal, $N_s(\theta)$, and the number of peaks due to pure noise, $N_n(\theta)$, as a function of a varying threshold $\theta$, and is empirically found to be

\begin{equation}
\label{eq:ThrObserv}
 \Phi(\theta) = \frac{N_s(\theta)}{\sqrt{N_s(\theta) + N_n(\theta)^2}}.
\end{equation}

\begin{figure}[htbp]
  \centering
  \includegraphics[width=1.16\columnwidth]{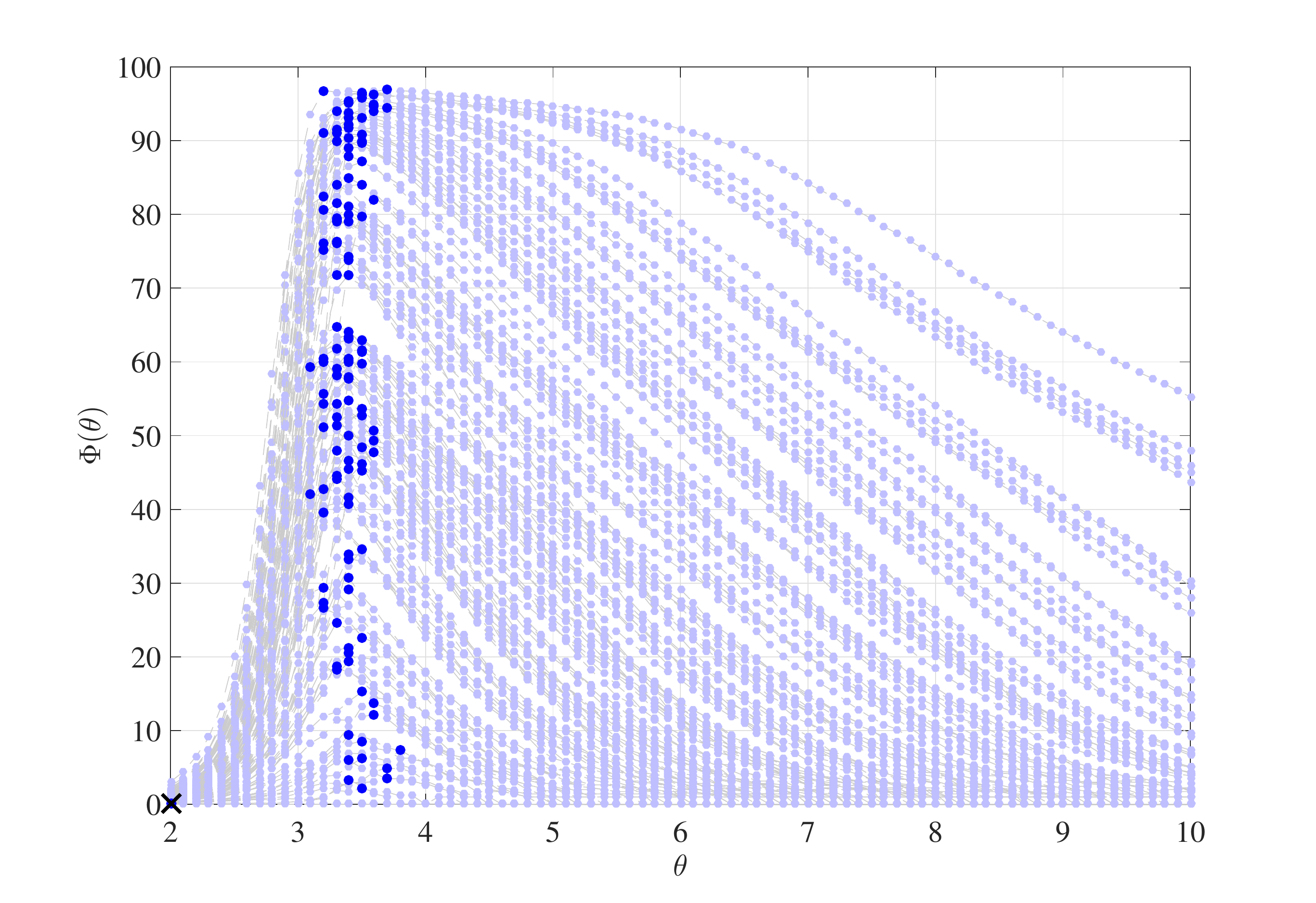} 
  \caption{(Color online) Observable $\Phi$ as a function of possible thresholds $\theta$ for 131 analysed frequency bands. The dark-blue dots correspond to the maximum values of $\Phi(\theta)$, while the black cross at $\theta=2$ corresponds to frequency bands where no signals could be identified, i.e. : [141,\,142], [166,\,167], and [190,\,191]~Hz (see details in Sec.~\ref{RecoveryEstim}).} 
   \label{fig:PhiVSsnr}
 \end{figure}

We identify frequency subbands, containing the CW signature, by using the procedure described in Sec.~\ref{Sec:GaussFilter}. We then select peaks with an amplitude $\mathcal{R}$ larger than $\theta$, ranging from 2~\footnote{The minimum value of $\theta$ is chosen to be slightly larger than the threshold set for initial peak selection $\Rth=\sqrt{2.5}\sim1.6$.} to 10 in steps of 0.1. Hence, we can estimate the number of peaks due to signal and noise in that band, able to surpass a given value of $\theta$, i.e., $N_{\mathrm{sn}}(\theta)$ [blue dots in Fig.~\ref{fig:TriangularFilterResp}~(a)], and the number of peaks due to pure noise in the same band, but assuming no signals were present, and surpassing the same $\theta$, i.e., $N_n(\theta)$ [red dots in Fig.~\ref{fig:TriangularFilterResp}~(a)].                 
Hence, we compute the number of peaks solely due to signal in a certain frequency band as $N_s(\theta) = N_{\mathrm{sn}}(\theta) - N_n(\theta)$. %, for different values of $\theta$. 
The observable $\Phi$, as a function of $ \theta$, is shown in Fig.~\ref{fig:PhiVSsnr}. 
 
As threshold in a given frequency subband we choose the optimal value, which is the one maximising $\Phi(\theta)$. As shown in Fig.~\ref{fig:PhiVSsnr}, this varies from a minimum of 3.1 to a maximum of 3.8, with an average of $\theta_{\mathrm{thr}} = 3.5$ for the bulk of frequency bands. 
In every subband, where a presumed CW signal has been identified, we select all peaks above this slightly varying threshold, and average the related frequencies corresponding to a same FFT midtime, obtaining at most $\mathcal{N}$ $(t_c, \bar{f}_p)$ pairs. 

A systematic study of how varies $\theta_{\mathrm{thr}}$ with varying $\Tobs$ and $\Tfft$ lies outside the scope of this paper as a value of $\theta_{\mathrm{thr}} = 3.5$ can be broadly adopted for further analyses, even on real detector data.
Such a value is indeed consistent with what it would be obtained by a visual inspection of the analysed frequency bands, and is however at the edge of what would be indistinguishable from Gaussian noise [as can be appreciated in Fig.~\ref{fig:TriangularFilterResp}~(a)].

%****************************************************************************************************
\subsection{\label{Sec:PeriodogramEstim} Spectral analysis of unequally spaced data}
%****************************************************************************************************

For all subbands where a putative signal has been identified, we face the problem to sift through the unevenly sampled data ($t_{c},\,\bar{f}_p$) in order to unveil any potential periodicity. In general, Fourier analysis is employed to characterise the frequency content of a signal, thus detecting possible periodicities. The FFT algorithm computes a Fourier decomposition under the assumption that the input data points are equally spaced in the time domain. However, approximate Fourier transform methods have also appeared in literature, which treat nonequispaced data (see, e.g.,~\cite{FEng2008}). More generally, there are other methods used to perform a spectral estimate of non-regularly spaced data, which are based on periodogram analysis.

The periodogram estimate is well suited to the problem of detecting a periodic signal in the presence of noise, and is the most common method to estimate the power spectrum of evenly and unevenly spaced data, due to the simplicity of its statistical behaviour~[see Eqs.~(4) and~(10) of~\cite{1982ApJ...263..835S} for evenly and unevenly spaced data, respectively].

%****************************************************************************************************
\subsubsection{\label{PPS} Peak power spectrum estimate and orbital period recovery}
%****************************************************************************************************
In the following we illustrate an alternative, and more straightforward, estimate of the power spectrum\footnote{We note that we use here ``power spectrum'' and ``periodogram'' reciprocally, although specifically the power spectrum is a theoretical quantity defined as an integral over continuous time, of which the periodogram is purely an estimate based on a amount of discrete data.}  with respect to the standard Lomb-Scargle periodogram estimator. We refer the reader to~\cite{1976Ap&SS..39..447L,1982ApJ...263..835S} for more details, including the statistical properties of the Lomb-Scargle periodogram. 

For every frequency subband (selected within every 1~Hz analysed band), where a signal has been pinpointed, we estimate the peak power spectrum as follows:

\begin{equation}
  \label{eq:peak_sp}
  \mathcal{S}(\nu_j) = \frac1{N} \, \left | \sum_{{k}=1}^{N} Y_{k} \, \mathrm{e}^{2  \pi \mathrm{i}  \nu_j  t_{c, {k}} } \right |^2,
\end{equation}
where $Y_{k}= \bar{f}_{p,\,k} - \left\langle \bar{f}_{p} \right\rangle_{N}$ are the deviations of $ \bar{f}_{p,\,k}$ from their mean value\footnote{The mean is a marginal statistic and is not important for judging periodicity. Hence, we can safely subtract the sample mean in Eq.~\eqref{eq:peak_sp} in order to obtain a series with zero mean. Not subtracting the sample mean would make the scale of the periodogram plot difficult to judge, mainly if $\mathcal{S}(0)$ is very large~\cite{ASHD}.}, and $ \bar{f}_{p,\,k}$ are the ${k}$th averaged peak frequencies. The number $N$ of averaged frequencies varies for every frequency subband, and is at most $N=\mathcal{N}$ when the signal contribution comes from all subpeakmaps.
We recall that the times $t_{c, {k}}$ correspond to the midtimes of the $k$th FFT (or, equivalently, subpeakmap) in the analysed frequency subband.

The number $N_{\nu}$ of frequency points $\nu_j$ (with $j = 1, \dots, N_{\nu}$) at which to compute the periodogram is given by
\begin{equation}
  \label{eq:PeriodgrFreq}
  N_{\nu} = \frac{\nu_{\mathrm{max}} - \nu_{\mathrm{min}}}{d\nu}  \,, 
\end{equation}
where $\nu_{\mathrm{min}} =0$~d$^{-1}$, $\nu_{\mathrm{max}} =10$~d$^{-1}$, and the step $d\nu = 1/(\Tobs \,r_\nu) $, with $r_\nu$ being a refinement factor. The values of $0$~d$^{-1}$ and $10$~d$^{-1}$ correspond to considering orbital periods from a few hours ($\sim 2.4$~h) up to infinity. This very wide range includes of course the source orbital periods we target [see Eq.~\eqref{eq:RandSpurParam}]. 

The resolution at which to evaluate the periodogram can be either refined or coarsened via $r_\nu$. A reasonable choice is a refinement in frequency resolution $r_\nu = 4$.
Hence, for the observation time $\Tobs = 1$~month considered here, we compute the periodogram for $1\,200$ frequencies. 

By taking the maximum value of $\mathcal{S}(\nu_j)$, and the inverse of the fundamental frequency, which is the frequency corresponding to $\max \mathcal{S}(\nu_j) $, i.e., $\nu^m$, we obtain the source orbital period, namely

\begin{equation}
  \label{eq:orbitalP}
  P = \frac1{\nu^m}\,. 
\end{equation}

The orbital periods estimated for 128 (out of 131) detected CW signals are depicted in Fig.~\ref{fig:AllParamEstim}~(c).

Appendix~\ref{Sec:PerPlot} illustrates an example of the periodogram appearance, and how the orbital eccentricity impacts on the number of harmonics. Perfect consistency with the Lomb-Scargle periodogram is also discussed.

%****************************************************************************************************
\subsection{\label{Sec:LSF} Least-squares fitting of sine waves and Parameter estimate}
%****************************************************************************************************
We illustrate here the method employed to estimate the signal frequency and all source orbital parameters but the orbital period. 

For every subband of interest, we perform a sinusoidal fit of the averaged peak frequencies $\bar{f}_p$:
\begin{equation}
  \label{eq:sinFit}
\bar{f}_{p,\,k} =  A_0 + \sum_{h=1}^{N_h=2} [A_{2 h-1} \cos(\Om \, h \, t_{c, {k}}) + A_{2 h} \sin(\Om \, h \, t_{c, {k}}) ]\,,
\end{equation}
with $k= 1,\, \dots,\, N$, $\Om = 2\,\pi/P$, and $P$ being the source orbital period recovered by the periodogram technique detailed in Sec.~\ref{PPS}. 
The $A$ coefficients are real numbers with dimension of the inverse of a time. 

In Appendix~\ref{spC} we provide details on the spectral content of the source orbital modulation, i.e. $\dot{R}(t)$, and we note (from Fig.~\ref{fig:NormHarAmpEcc}) that only two harmonics are necessary to describe the spectral content of $\dot R/c$.
Hence we chose $N_h=2$. 

When then have to solve the overdetermined linear system of $N>N_h$ equations, given by Eqs.~\eqref{eq:sinFit}, in the unknowns $A_{0,\,1,\,2,\, 3,\, 4}$, i.e.:
\begin{equation}
\label{eq:LinSisLeastSqMt}
\mathcal{A}  \cdot X = \mathcal{Y},
\end{equation}

with 
\begin{equation}\label{eq:CosSinMatrix}
\small
\mathcal{A} =
\begin{bmatrix}
1 &  \cos(\Om \, t_{c, 1}) &  \sin(\Om \, t_{c, 1}) &  \cos(2\,\Om \, t_{c, 1}) &  \sin(2\,\Om \, t_{c, 1}) \\
1 & \cos(\Om \, t_{c, {2}}) &  \sin(\Om \, t_{c, {2}}) &  \cos(2\,\Om \, t_{c, {2}}) &  \sin(2\,\Om \, t_{c, {2}}) \\
. & . & . &  . &  . \\
. & . & . &  . &  . \\
. & . & . &  . &  . \\
1 & \cos(\Om \, t_{c, {N}}) &  \sin(\Om \, t_{c, {N}}) &  \cos(2\,\Om \, t_{c, {N}}) &  \sin(2\,\Om \, t_{c, {N}}) 
 \end{bmatrix},
\end{equation}
\begin{equation}	
X=
  \begin{bmatrix}
    A_0\\A_1\\A_2\\A_3\\A_4
  \end{bmatrix},
\quad
 \mathcal{Y} =
  \begin{bmatrix}
    \bar{f}_{p,\,1}\\\bar{f}_{p,\,2}\\. \\. \\. \\\bar{f}_{p,\,N}\\
  \end{bmatrix}.
\end{equation}	
The exact solution is obtained by using the \emph{least-squares method}.

Equation~\eqref{eq:sinFit} is an equality between $\bar{f}_{p,\,k}$ and a sum of sinusoids at frequencies $1/P,\,2/P$, with $1/P$ the first harmonic (i.e., the fundamental frequency). Technical details on how establishing the existence of these two peak power spectrum harmonics can be found in Appendix~\ref{Sec:SecHarmPeakSpec}.
The amplitudes of the first and second sinusoids in Eq.~\eqref{eq:sinFit} are given by $\mathcal{H}_1 = \sqrt{A_1^2 + A_2^2}$ and $\mathcal{H}_2 = \sqrt{A_3^2 + A_4^2}$, respectively, while $A_0$ is the so-called DC value. This term corresponds to the amplitude of a cosine wave with zero frequency [note, in fact, the presence of ones in Eq.~\eqref{eq:CosSinMatrix}].
		
In order to solve Eq.~\eqref{eq:LinSisLeastSqMt}, we multiply each member of such an equation by $\mathcal{A}^{\mathrm{T}}$, with the superscript $T$ denoting the transpose matrix, and obtain
\begin{equation}
\label{eq:BettLinSisLeastSqMt}
\mathcal{B}  \cdot X = \mathcal{D},
\end{equation}
where $ \mathcal{D} = \mathcal{A}^{\mathrm{T}} \cdot \mathcal{Y}$, and $\mathcal{B} = \mathcal{A}^{\mathrm{T}}\cdot \mathcal{A}$ is a ($5 \times 5$) square matrix. The number $5$ comes from considering the $b=2\,N_h+1$ amplitudes $A_{0\,,1\,, 2\,,3\,,4}$ (chosen $N_h=2$).

The estimate of the unknown $X$ parameters is obtained by solving Eq.~\eqref{eq:BettLinSisLeastSqMt}, i.e.,
\begin{equation}
\label{eq:UnX}
X= \mathcal{B}^{-1} \cdot \mathcal{D}\,;
\end{equation}
$X$ is the least-squares solution that minimises the $(N - 5)$ degrees-of-freedom $\mathcal{\chi}^2$ variable
\begin{equation}
\label{eq:chi2}
\mathcal{\chi}^2 = \sum_{k=1}^{N} (\bar{f}_{p,\,k} - \mathcal{A}\cdot X)^2,
\end{equation}
where $ (\bar{f}_{p,\,k} - \mathcal{A}\cdot X)$ are the residuals, whose mean indicates the accuracy of the solution found: the closer to 0, the more accurate $X$. 
The inverse matrix $\mathcal{C} =  \mathcal{B}^{-1}$ is the \emph{covariance matrix} of the $A_{0\,,1\,, 2\,,3\,,4}$ parameters, and the elements on the diagonal of $\mathcal{C}$ are proportional to the variance of $A_{0\,,1\,, 2\,,3\,,4}$. Hence, their uncertainties are
\begin{equation}
\label{eq:uncertParam}
dA_b = \sqrt{\mathcal{C}_{bb}}\,\frac{1}{\Tfft}\,,\quad b=0,\,\dots\,,4.
\end{equation}

Having then solved Eq.~\eqref{eq:UnX}, we can estimate the $A_{0\,,1\,, 2\,,3\,,4}$ parameters, and explicitly write Eq.~\eqref{eq:sinFit} for every $k$ as
\begin{widetext}   
\begin{equation}
\label{eq:sinFit2harm}
%\[
\bar{f}_{p,\,k} =  \highlight{A_0} + [\LightHighlight{A_{1} \cos(\Om \, t_{c, {k}}) + A_{2} \sin(\Om \, t_{c, {k}})} + \lowHighlight{A_{3} \cos(2\Om \, t_{c, {k}}) + A_{4} \sin(2\Om \, t_{c, {k}})} ],
%\]
\end{equation}
\end{widetext}
which we can compare against the received frequencies, which are modulated due to the source orbital motion, i.e. :
\begin{widetext} 
\begin{equation}
\label{eq:sinFit2harmRdot}
f_m= f - f\,\frac{\dot R(t_{c, {k}})}{c} =  \highlight{f} -
 f\, \asini  \Om \,[ \LightHighlight{\cos(\Om (t_{c, {k}}-\tAsc))} + \lowHighlight{\kappa \cos(2\Om (t_{c, {k}}-\tAsc)) + \eta \sin(2\Om (t_{c, {k}}-\tAsc))} ],
\end{equation}
\end{widetext}
with the orbital Doppler modulation contribution obtained by deriving Eq.~\eqref{eq:RoemDsmallEcc} with respect to time.

From the comparison of the dark-gray highlighted terms of Eqs.~\eqref{eq:sinFit2harm} and~\eqref{eq:sinFit2harmRdot} we find the signal frequency
\begin{equation}
\label{eq:freqRec}
f = A_0\,.
\end{equation}

By assuming to perform a fit with a sinusoidal function, i.e., by comparing the gray highlighted terms of Eqs.~\eqref{eq:sinFit2harm} and~\eqref{eq:sinFit2harmRdot}, we obtain an estimate of the projected orbital semi-major axis
\begin{equation}
\label{eq:apRec}
\asini = \frac{\mathcal{H}_1}{A_0  \, \Om}\,.
\end{equation}

Finally, by comparing the light-gray highlighted terms, we find the orbital eccentricity given by the ratio of the two harmonic amplitudes:
\begin{equation}
\label{eq:eccRec}
e = \frac{\mathcal{H}_2}{\mathcal{H}_1}.
\end{equation}
In Appendix~\ref{spC} we show that, for very low eccentricities, only the first harmonic of $\dot{R}(t)/c$ exists (see Fig.~\ref{fig:NormHarAmpEcc}).
Hence, for these cases no eccentricity estimates can be provided (as we would need the second harmonic as well).

The argument of periapse is given by
\begin{equation}
\label{eq:PerRec}
\argp = \arctan\left(\frac{A_4}{A_3}\right),
\end{equation}
where we used that $e=\sqrt{\kappa^2+\eta^2}$ and $\argp = \arctan (\eta/\kappa)$, due to Eqs.~\eqref{eq:kappa} and~\eqref{eq:eta}.

Lastly, we can compute the time of periapse from Eq.~\eqref{eq:28}:

\begin{equation}
\label{eq:tpRec}
\tPeri = \frac1{\Om} \left[\argp - \arctan\left( \frac{A_2}{A_1}\right) \right] \,, 
\end{equation} 
being 
\begin{equation}
\label{eq:tascRec}
\tAsc  = - \frac1{\Om} \arctan\left( \frac{A_2}{A_1} \right).
\end{equation}

%****************************************************************************************************
\section{\label{RecoveryEstim} Results: detection and parameter estimation}
%****************************************************************************************************

We claim 128 signal detections out of total 131. The frequency bands where our algorithm fails to detect a signal are [141,\,142]~Hz, [166,\,167], and [190,\,191]~Hz. The gravitational-wave strain amplitudes of the three not-detected signals is $h_0\sim 10^{-24}$ for one month of single-detector data set we process. The signals are missed as they are not loud enough to stand out of the background noise, due to either a large orbital Doppler modulation or modulated signal patterns with too much asymmetric horns [depending on the orbital parameters, see Eq.~\eqref{eq:modDepth}]. 

\begin{figure*}[htbp] 

  \raggedright (a)\hspace*{\columnwidth}(b)\\[-0.03cm]
  \includegraphics[clip,width=\columnwidth]{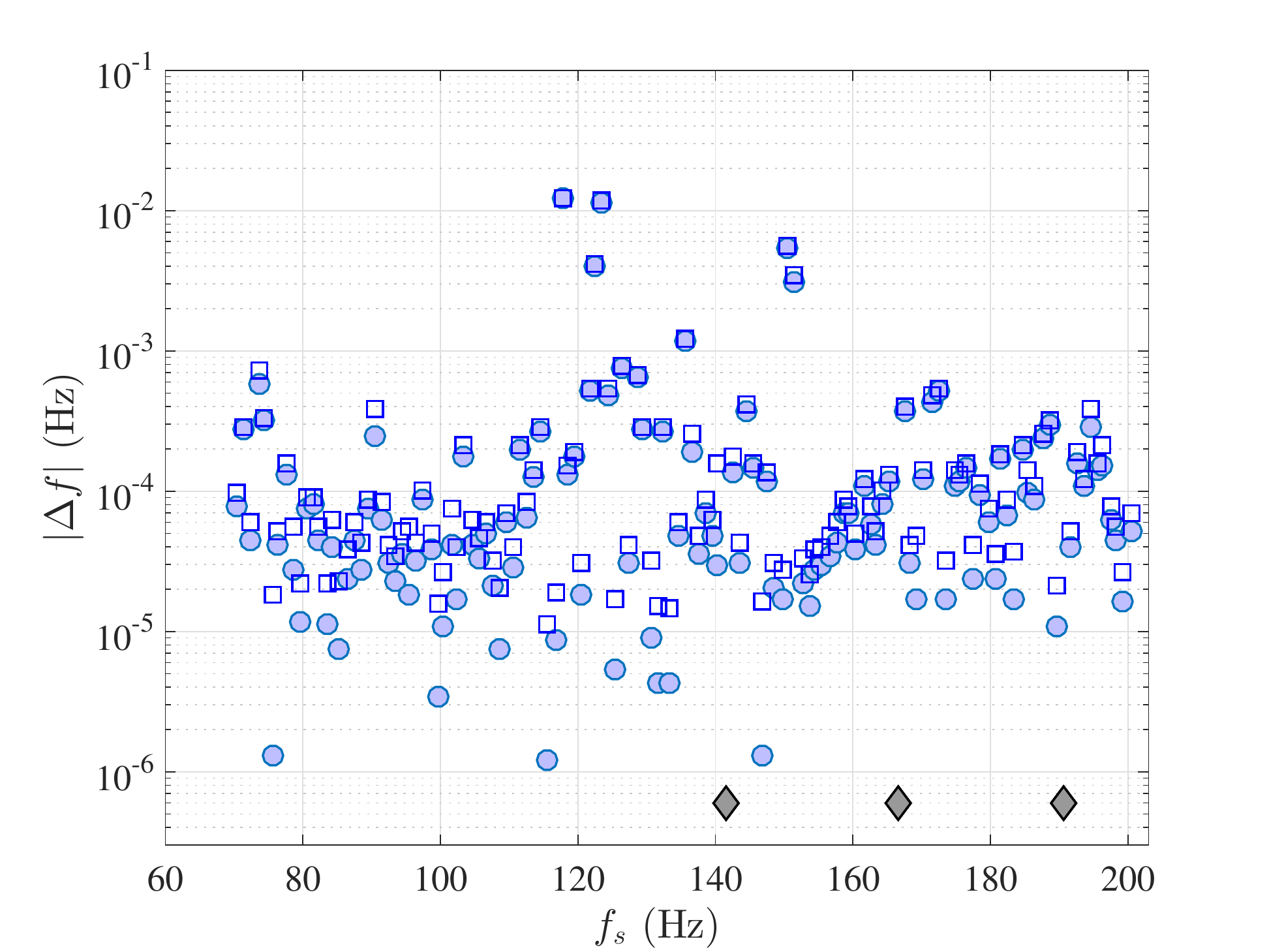}
  \includegraphics[clip,width=\columnwidth]{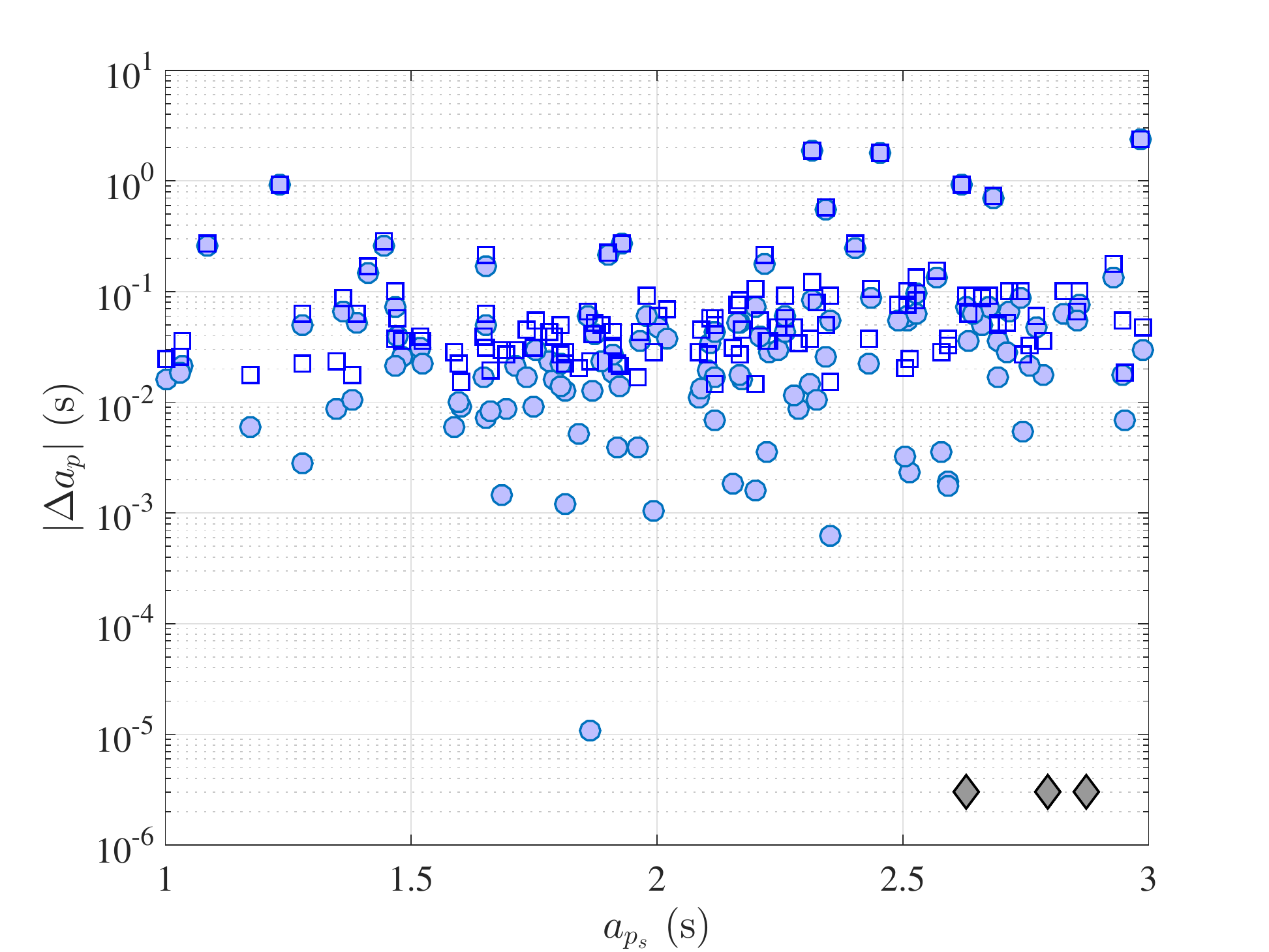}\\[-0.03cm]

  \raggedright (c)\hspace*{\columnwidth}(d)\\[-0.03cm]
  \includegraphics[clip,width=\columnwidth]{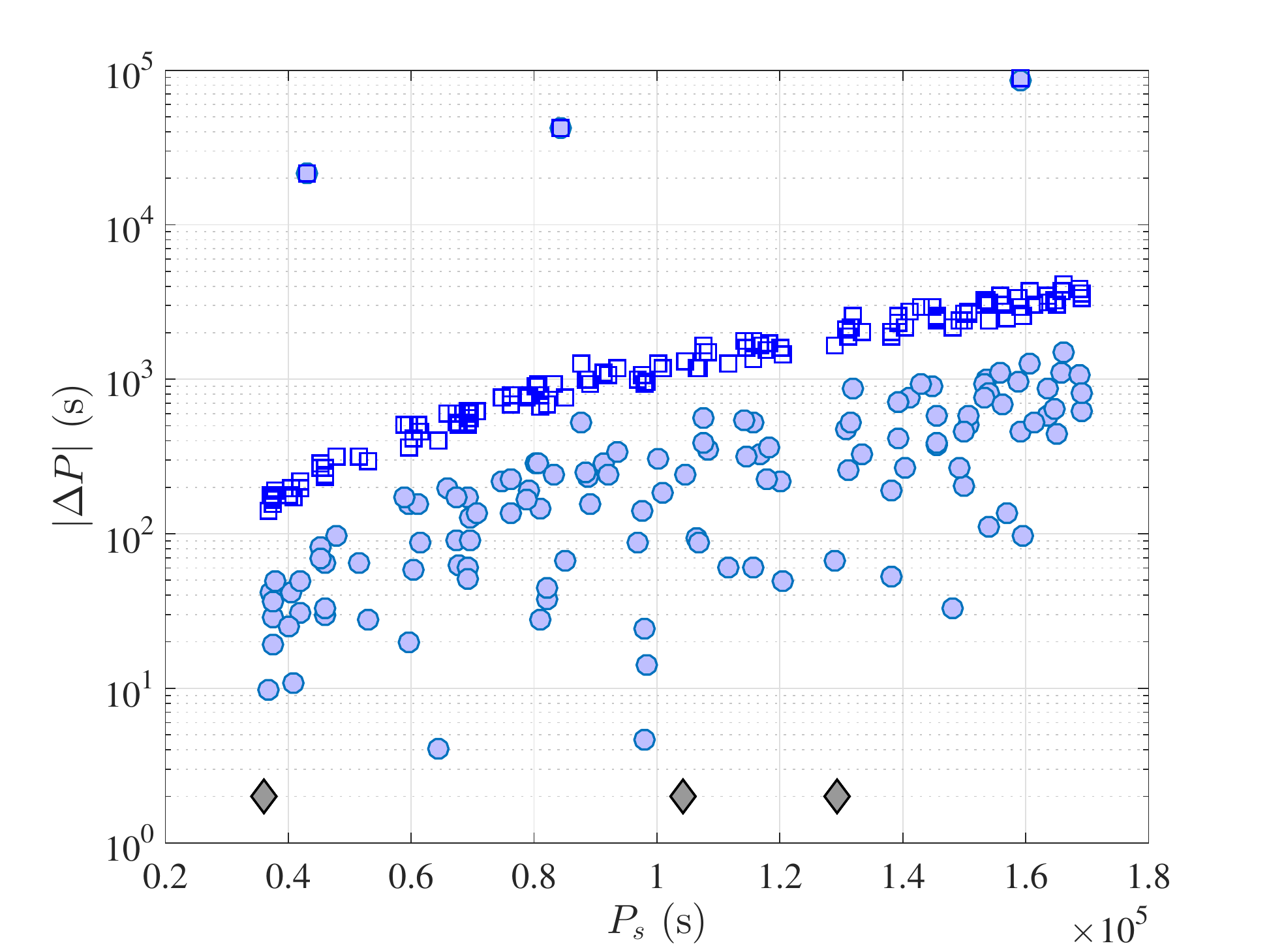}
  \includegraphics[clip,width=\columnwidth]{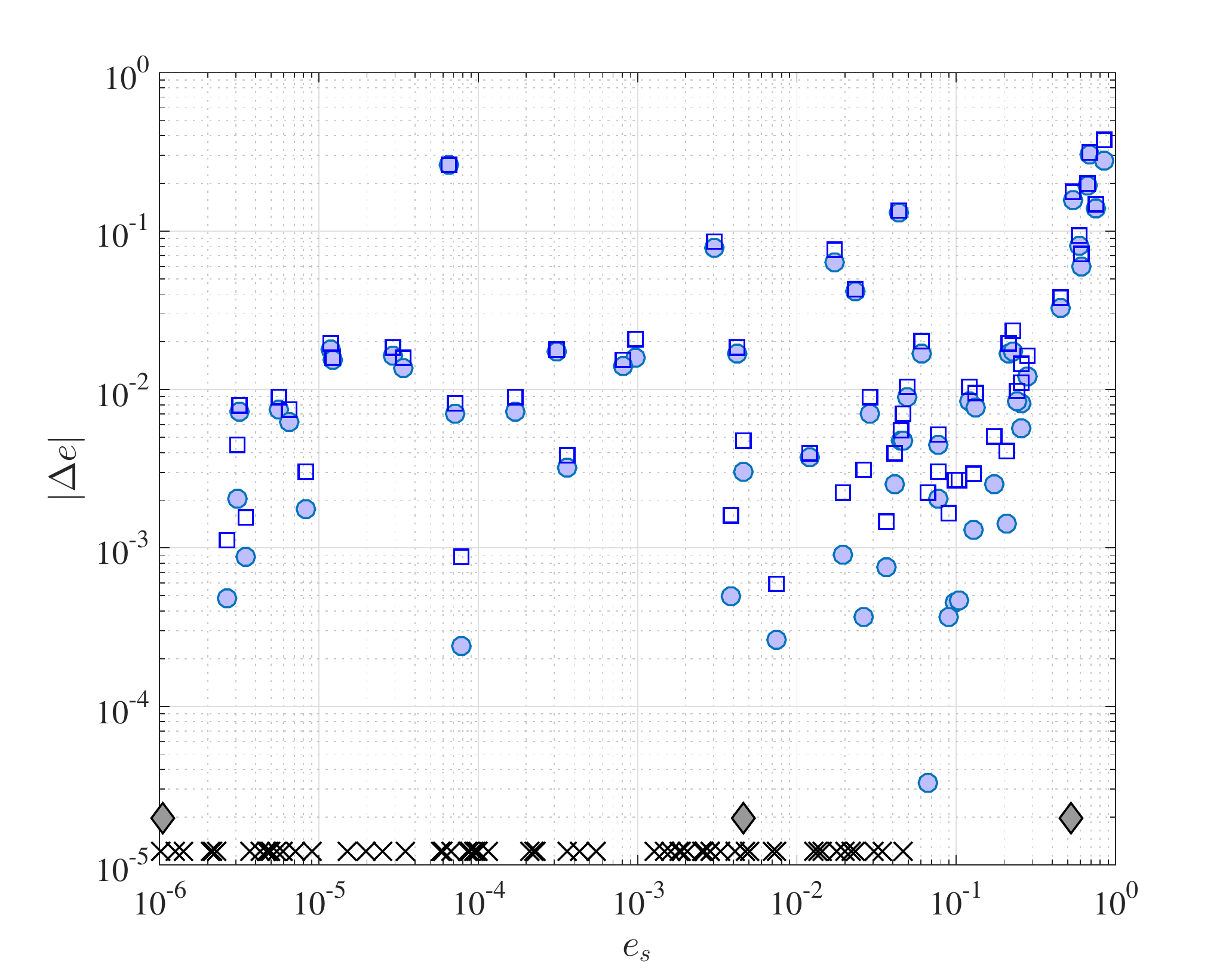}\\[-0.3cm]

  \raggedright (e)\hspace*{\columnwidth}(f)\\[-0.03cm]
  \includegraphics[clip,width=\columnwidth]{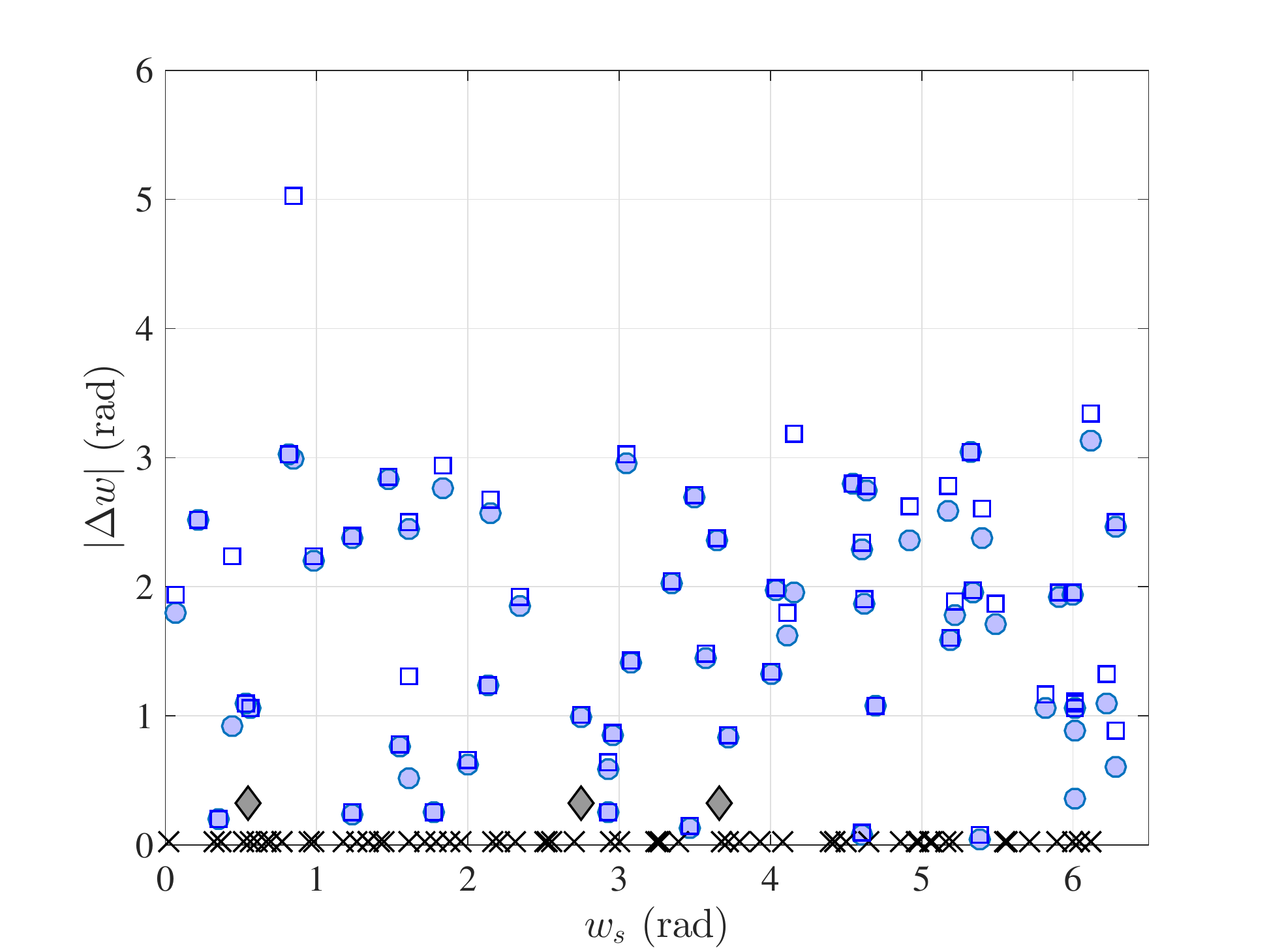}
  \includegraphics[clip,width=\columnwidth]{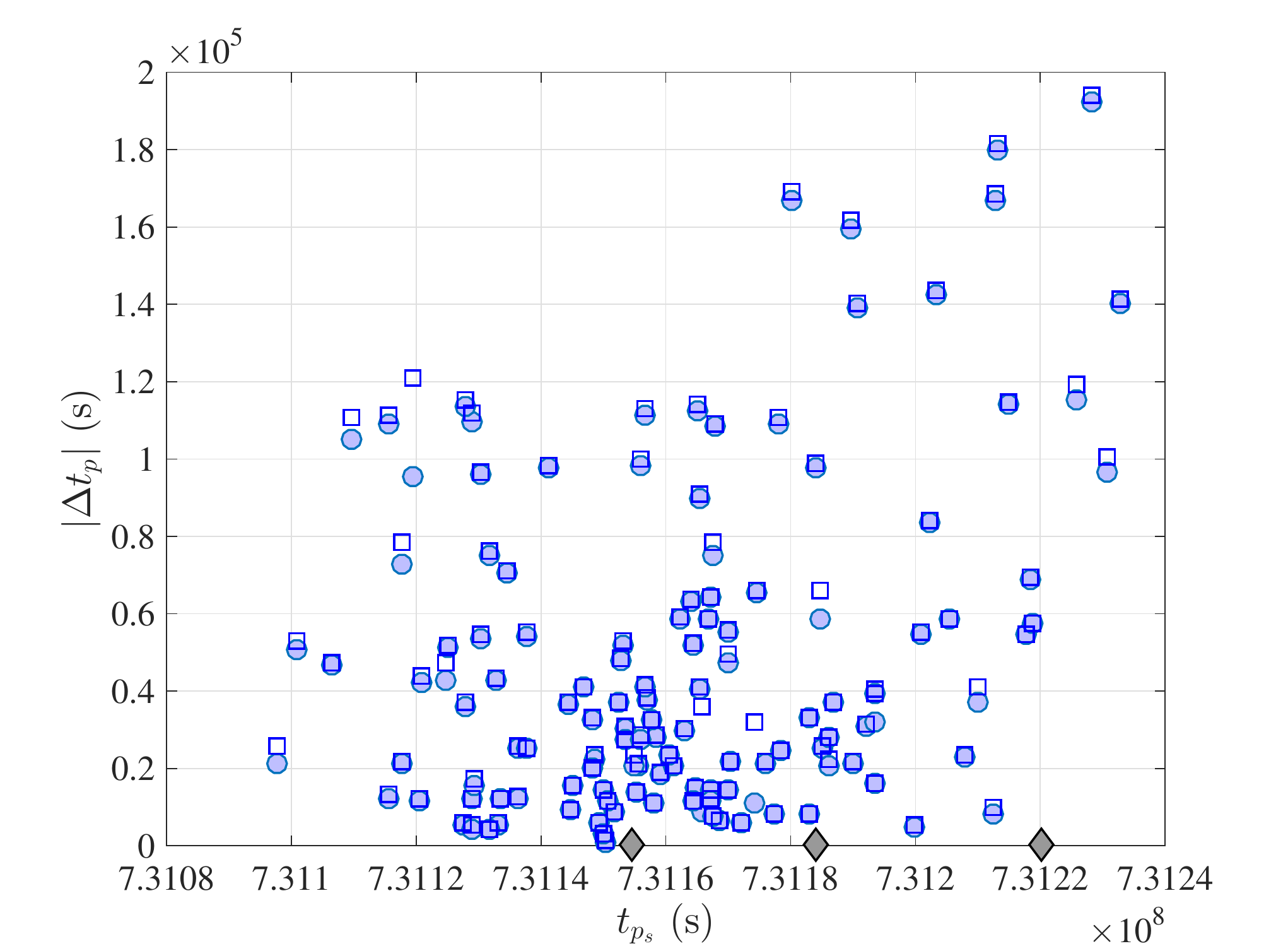}\\[-0.3cm]
  \caption{(Color online) Parameter estimation for detected signals as a function of 131 injected signal (a) frequency $f_s$, (b) projected orbital semi-major axis $a_{p_{s}}$, (c) orbital period $P_s$, (d) eccentricity $e_s$, (e) argument of periapse $\omega_s$, and (f) time of periapse passage $t_{p_{s}}$. The light-blue filled circles show the absolute value of the difference between the injected (true) and recovered values of signal parameters. The empty squares show the quoted one-sigma errorbars corresponding to each estimate. The filled diamonds mark the parameters of signals that have not been detected (and for which no parameter estimates can be provided). The crosses in panels (d) and (e) indicate the parameters of 69 signals for which no estimate can be obtained as the second harmonic is not found in the corresponding periodogram. Hence, the estimate is provided only for 59 signals.} 
 \label{fig:AllParamEstim}
\end{figure*}

Figure~\ref{fig:AllParamEstim} shows the parameter estimates for 128 detected signals; the parameters of the three signals that could not be detected are denoted by filled diamonds in all panels.
The offsets on the vertical axes give the absolute value of the difference between estimated and true signal parameters (denoted with $r$ and $s$ subscripts, respectively):
\begin{equation}
\label{eq:GenParam}
\Delta \mathcal{O} = \mathcal{O}_r - \mathcal{O}_s,
\end{equation}
where $\mathcal{O}$ refers to a generic parameter among those shown in Fig.~\ref{fig:AllParamEstim}. 

We observe that the recovery of signal frequency $f_s$, projected orbital semi-major axis $a_{p_{s}}$ and orbital period $P_s$ is generally quite good. The behaviour of the one-sigma orbital period uncertainty $dP$ [open squares in panel (c)] is larger than typical actual offsets, and reflects the scaling of $dP$ with $P^2$, i.e.:
\begin{equation}
  dP = P^2\, d\nu,
 \end{equation}
 with $d\nu = (4\,\Tobs)^{-1}$, as described in Sec.~\ref{PPS}. Obviously, the higher the orbital period and the less precise its estimate, as the number of orbits observed during $\Tobs$ decreases.
 The offsets larger than $10^4$~s in Fig.~\ref{fig:AllParamEstim}~(c) correspond to three signals where the periodogram estimate does return an orbital period which is half of the true value. This happens for sources with an orbital period close to multiples and submultiples of the Earth's sidereal day, especially if they are not strong enough to stand out of noise, and have a quite high eccentricity. In these circumstances, multiples and submultiples of the Earth periodicity can be mistaken for the signal periodicity. 
A way to bypass this contingency consists of performing a sinusoidal fit with two harmonics (similarly to what discussed in Sec.~\ref{Sec:LSF}), looping over several target orbital periods, and then choosing the orbital period value for which the fit exhibits the smallest residuals. Such approach might be computationally demanding, especially if we consider a large number of orbital periods. This technique will be however used in a separate study aimed at improving source parameter estimation.
 
The uncertainties on the source parameters are obtained by using standard error propagation rules for uncorrelated variables from Eqs.~\eqref{eq:freqRec},~\eqref{eq:apRec},~\eqref{eq:eccRec},~\eqref{eq:PerRec},~\eqref{eq:tpRec}, and using as $dA_b$ uncertainties (with $b=0,\dots,4$) those given by Eq.~\eqref{eq:uncertParam}.
 
 The estimate of signal eccentricity $e_s$, argument of periapse $\omega_s$, and time of periapse passage $t_{p_{s}}$, is more delicate. The crosses in panels (d) and (e) of Fig.~\ref{fig:AllParamEstim} correspond to 69 cases where a signal is detected but no estimate can be provided for $e_s$ and $\omega_s$, as the second harmonic is not found in the corresponding periodogram, and hence the recovered eccentricity would be indistinguishable from zero (as also described in Appendix~\ref{spC}). 
 As a reminder, we used the criterion discussed in Appendix~\ref{Sec:SecHarmPeakSpec} to check if the second harmonic of the periodogram exists. If the second harmonic does not exist, the current method does not have the ability to estimate the eccentricity. Further studies are needed to understand what strategy to use for such situations.
 
 The $t_{p_{s}}$ estimate is generally poor, and would improve with more precise estimates of $\omega_s$ and $P_s$.
  
We highlight that in Eq.~\eqref{eq:sinFit2harmRdot} we use the small-eccentricity approximation for $\dot R/c$, which works overall well to estimate also high eccentricities. Future studies will be however devoted to understand how to make suitable use of Eq.~\eqref{OrDoppMod} in a least-squares fitting approach.

By using the novel method presented here we are able to detect the bulk ($\sim 97\%$) of injected signals in a paltry amount of time (see Sec.~\ref{Sec:TeoCtot}), and to perform satisfactory parameter estimates. All our estimates can be however improved by using a hierarchical approach, which use the current parameter estimates to obtain an approximate demodulation of the source orbital motion, and then produce longer FFTs, by assuming such a partially known modulation. Then, the same procedure can be reiterated on the new data set, thus improving the parameter estimation and search sensitivity by a factor depending on the square root of how much the FFT duration can be increased. Such approach will be however implemented in a separate pipeline. 

This is the first algorithm in literature able to provide estimates for orbital period, orbital eccentricity and argument of periapse. 

%****************************************************************************************************
\section{\label{Sec:h0recovery} Recovery of signal strain amplitude}
%****************************************************************************************************

The current search is sensitive to a combination of $h_0$ and $\cos \iota$, which is given by~\cite{VelaSpinDAbadie:2011md}
\begin{equation}
\label{eq:H0andh0}
H_0 = \frac{h_0}{2}\, \sqrt{1 + 6\,\cos^2\iota +\cos^4\iota}.
\end{equation}

\begin{figure}
  \centering
  \includegraphics[width=1.01\columnwidth]{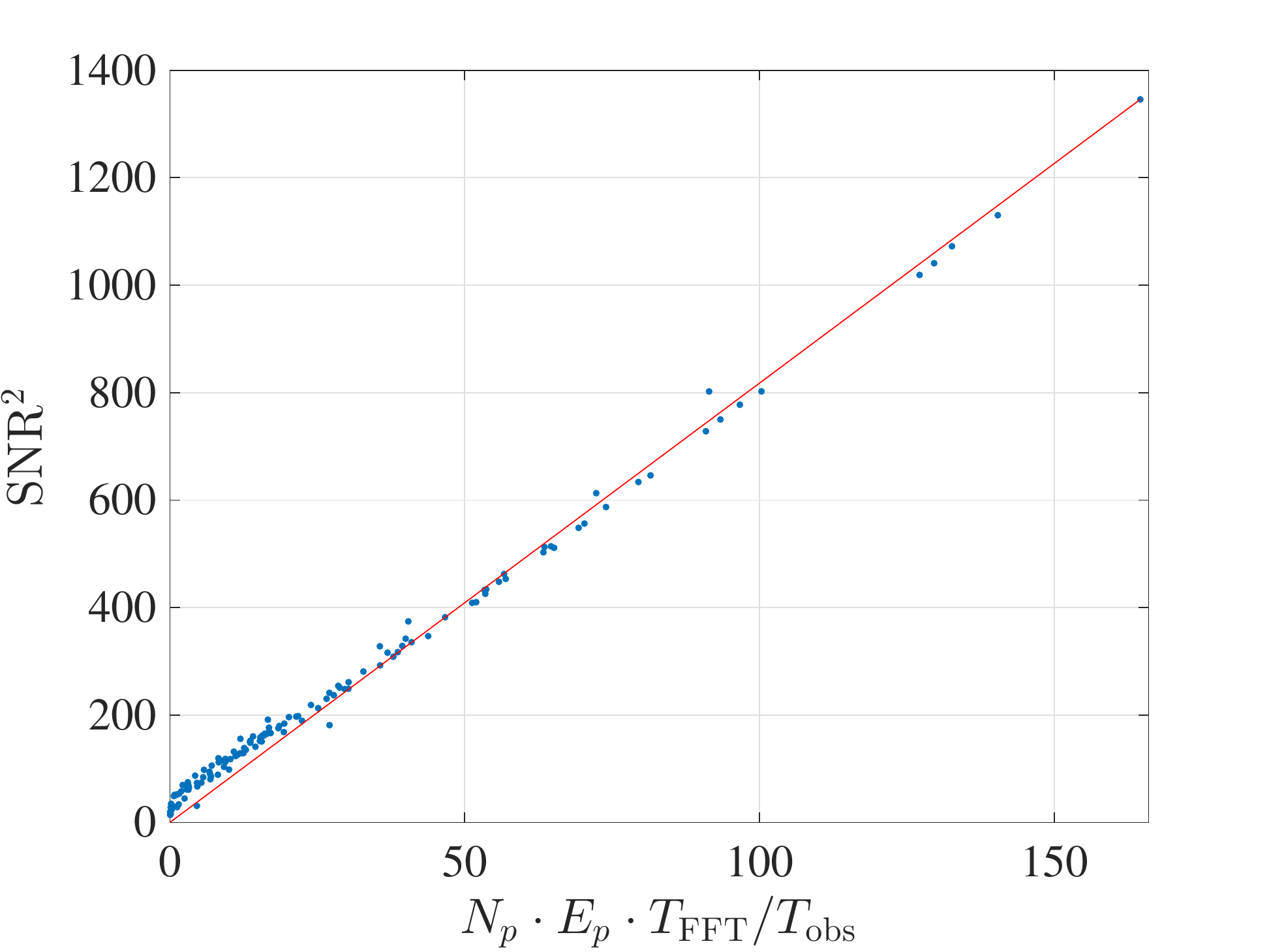}
  \caption{(Color online) Observed $\mathrm{SNR}^2$ versus $N_p \, E_p \, \Tfft/\Tobs$ (points), and related linear fit (straight line). } 
  \label{fig:LinFit}
\end{figure}

\begin{figure}  
  \includegraphics[clip,width=\columnwidth]{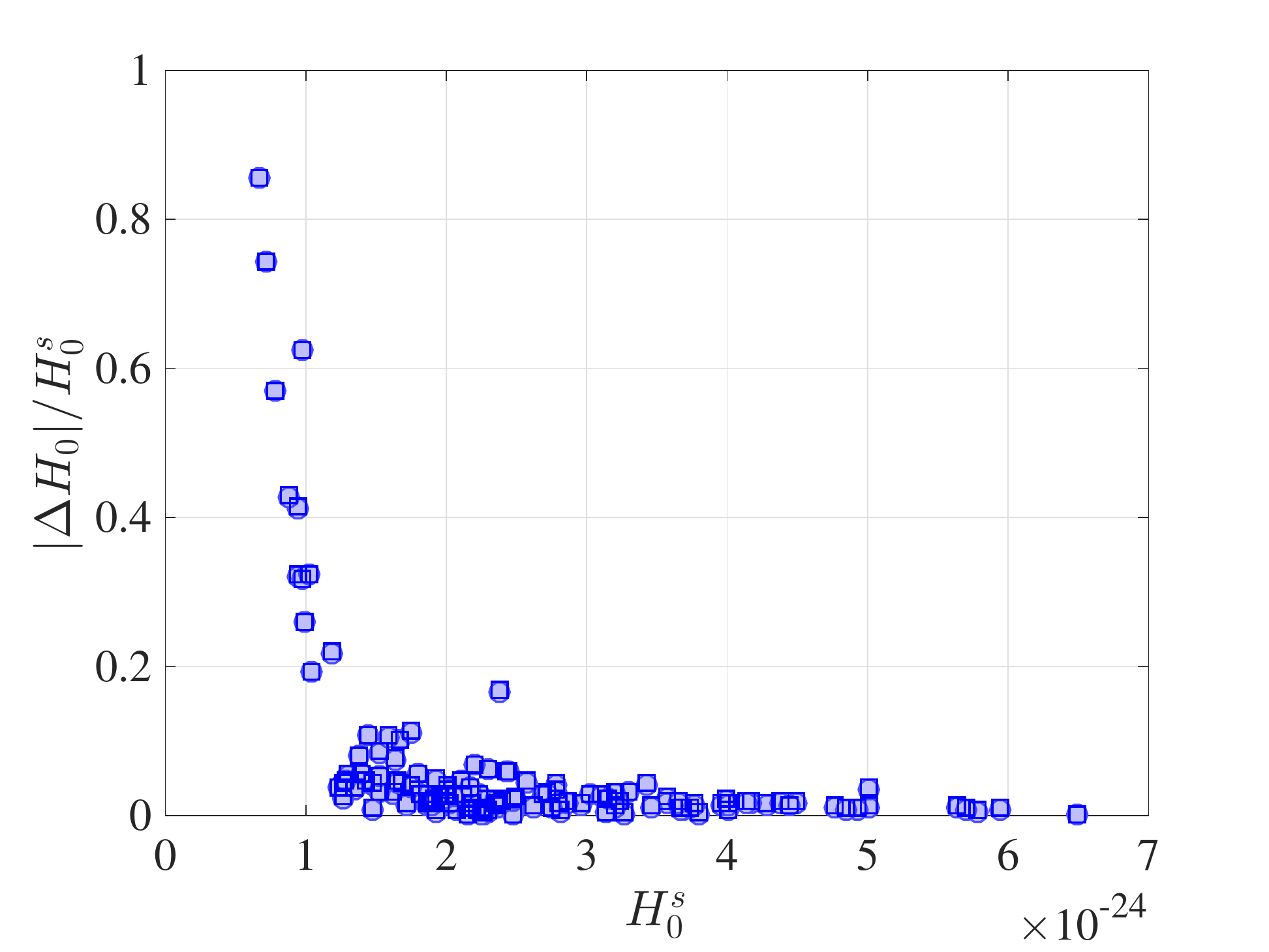}
  \includegraphics[clip,width=\columnwidth]{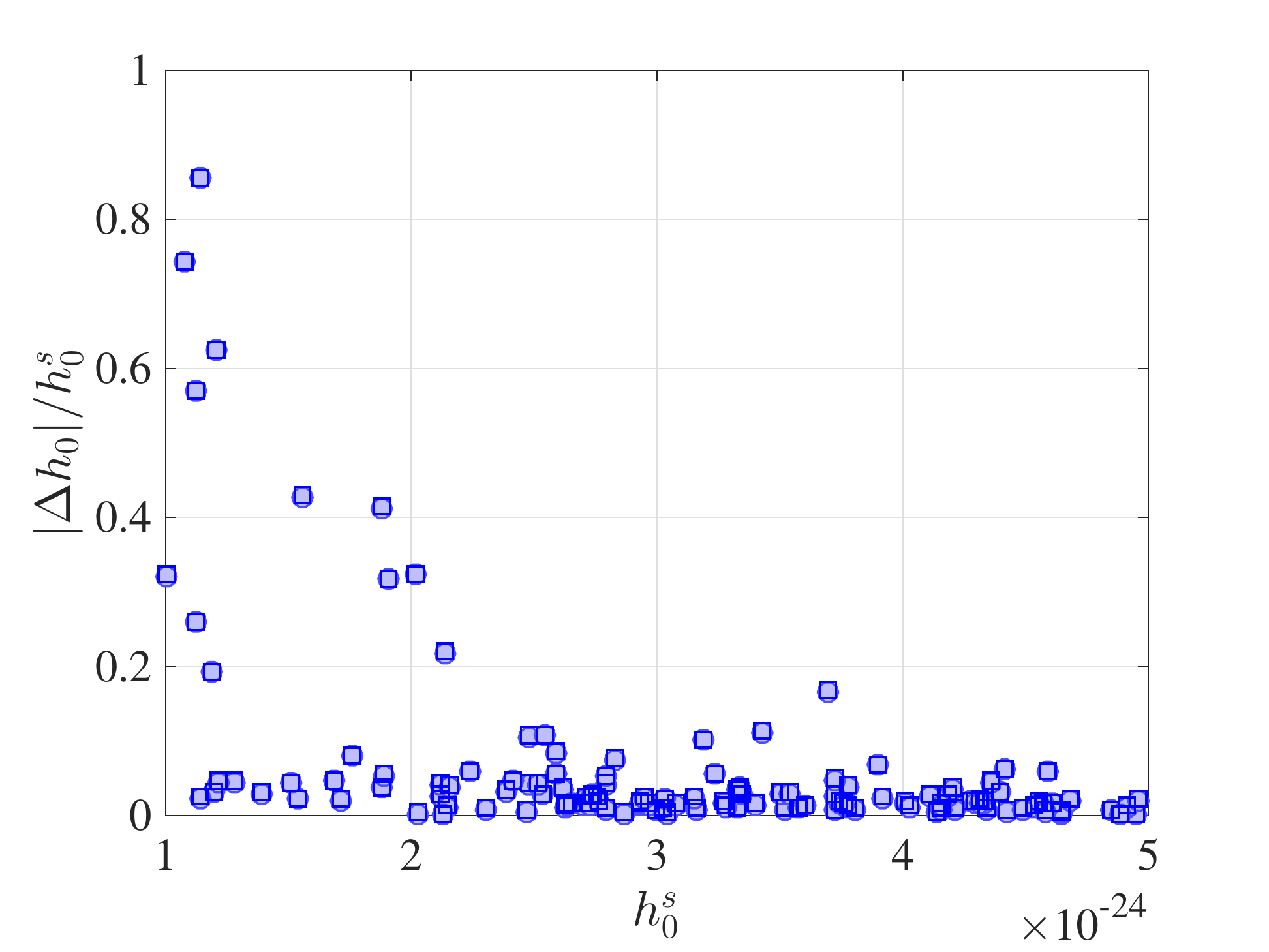}\\[-0.03cm]
  \caption{(Color online) Fractional errors in strain amplitude estimates represented by blue filled circles, and almost superimposed there are the one-sigma error bars divided by the true values. The estimates are provided for 128 out of 131 original inspected frequency bands, where an artificial CW signal has been injected and detected (in each of them).}
  \label{fig:hHrec}
\end{figure}

We can define a theoretical SNR in terms of the $H_0$ amplitude:
\begin{equation}
\label{eq:SNRandH0}
\mathrm{SNR} = H_0\,\sqrt{\frac{\Tfft}{S_h}},
\end{equation}
and find an empirical relation between such SNR and the number of peaks selected above threshold, on average, per FFT in a frequency band where a signal has been identified, and that are properly weighed by taking into account the signal amplitude, i.e.:
\begin{equation}
\label{eq:hoEstim}
\mathrm{SNR}^2 = N_p \, E_p \, \frac{\Tfft}{\Tobs} + \mathrm{corr}.
\end{equation}
We labelled $N_p$ as the number of peaks in a small frequency band where a CW signal has been found, and being above the threshold established in Sec.\ref{Sec:ThresholdChoicePA}, while $E_p$ is the average of the squared peak amplitude above threshold (i.e., $\mathcal{R} > \theta_\mathrm{thr}$) in the same band. The frequency band of the modulated signal pattern is 
identified according to what outlined in Sec.~\ref{Sec:GaussFilter}. 
Figure~\ref{fig:LinFit} shows the relation expressed by Eq.~\eqref{eq:hoEstim} [with the SNR given by Eq.~\eqref{eq:SNRandH0}], which is linear apart from a correction factor $\mathrm{corr}$, necessary to account for the discrepancies obtained for low SNR signals. We note that this correction factor is included in the uncertainty estimates provided in Fig.~\ref{fig:hHrec}.

In the top (bottom) panel of Fig.~\ref{fig:hHrec} we show the consistency of the estimated signal amplitudes $H_0^r$ ($h_0^r$) with the true values $H_0^s$ ($h_0^s$), together with the relative uncertainties that are especially small.

We emphasize that such a method can be used to estimate the gravitational-wave strain amplitude also in CW searches for isolated NSs, as there is no dependency on orbital parameters. Hence, by using Eqs.~\eqref{eq:SNRandH0} and~\eqref{eq:hoEstim}, and interpolation or extrapolation (if needed) schemes, for a given set of $N_p$, $E_p$, $\Tfft$ and $\Tobs$, we can find the corresponding $H_0$ of a signal, which has been previously identified in a frequency band.

%****************************************************************************************************
\section{\label{Sec:SensEstim} Sensitivity estimate}
%****************************************************************************************************

The customary \textit{modus operandi} to rigorously estimate the sensitivity of a search for CW signals is based on cumbersome Monte-Carlo simulations, which require consecutive signal injections.
We circumvent such expensive approach, opting for a less accurate, but expeditious, iterative procedure that provides however reliable estimates.

\begin{figure}[htbp] 
  \includegraphics[clip,width=1.1\columnwidth]{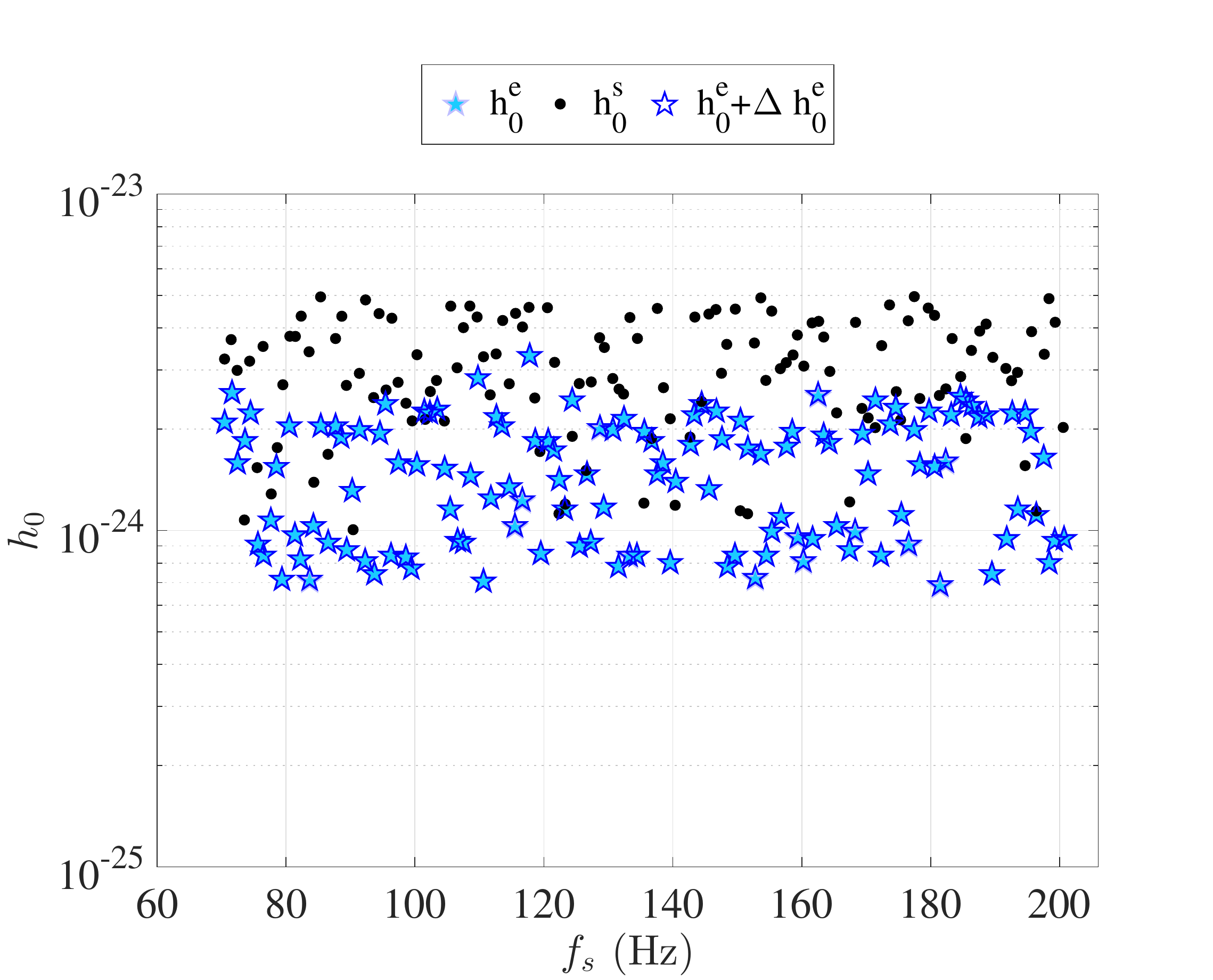} 
  \caption{(Color online) Detectable strain amplitudes versus $128$ signal frequencies. The dots are the injected strain values, the filled stars are the minimum strain amplitudes
  that can be detected (over one month of single-detector data), and the (hardly visible) empty stars the one-sigma errorbars associated to these estimates.}
  \label{fig:SensEstimh}
\end{figure}

In every 1-Hz frequency band, where a CW signal has been found, we decrease the peak amplitude $\mathcal{R}$ by a factor varying from 1 up to 0.1, in steps of 0.1, and select all peaks above the initial threshold $\Rth=\sqrt{2.5}$. To these rescaled data we apply, 
recursively, the cascade of four filters introduced in Sec.~\ref{Sec:GaussFilter} until a CW signal can no longer be detected.
This translates into testing essentially the performance of the Gaussian like filter, which has the final say-so to consider a signal detected or not detected.
Such approach is equivalent to keep fixed all parameters of the sources injected in every 1-Hz band (as discussed in Sec.~\ref{Sec:SignParam}), but the strain amplitude $h_0^s$, which decreases by a factor that can be obtained from Eq.~\eqref{eq:hoEstim}. We determine thus $N_p$ and $E_p$, based on the scaled peak amplitude $\mathcal{R}$, and crossing the same threshold settled in Sec.~\ref{Sec:ThresholdChoicePA}. Hence, we can estimate the minimum strain amplitude $h_0^e$ that can be detected by the filter cascade. The resulting detectable strain amplitude $h_0^e$ is plotted in Fig.~\ref{fig:SensEstimh} against signal frequency for 128 sources, as we excluded the three 1-Hz frequency bands where no signal has been previously detected.

The results obtained in Fig.~\ref{fig:SensEstimh} are promising, mainly considering they derive from one month of single-detector data, albeit with the caveat of being in Gaussian-noise. 
The reason of the broad variability in terms of sensitivity estimation must be attributed to the performance of the Gaussian like filter, which needs to be fine-tuned and enhanced. This is part however of a supplementary study, which aims also at carrying investigations over to real interferometer data, with the goal of both testing and strengthen the performance of the Gaussian like filter, and provide more accurate sensitivity estimations. 
In the circumstance of real detector data we expect several outliers to compromise the ability of such a filter to identify putative CW signals. We plan however to resort to follow-up studies, and coincidence-based methods, to verify the presence of a given signal. 

%****************************************************************************************************
\section{\label{Sec:TeoCtot} Computing cost model}
%****************************************************************************************************

The computational cost of the machinery here presented is estimated to be around 2.4 CPU hours on a 1.4 GHz Intel Core i5 processor. This estimate is based on timing the different pieces of the analysis, without counting the SFDB production, which is discussed in \cite{VCompMod}.

The total computing cost $C_{\mathrm{tot}}$ needed to both claim a detection and estimate source parameters is
\begin{equation}
\label{eq:Ctot}
C_{\mathrm{tot}}= C_1 + C_2,
\end{equation}
where $C_1$ is a computing cost scaling linearly with the time length (i.e., the observation time) of the data being analysed, with the number of inspected frequency bands $N_b$, and inversely with the FFT duration. Furthermore, if we would analyze data from different detectors independently, $C_1$ would scale also with the number of interferometers considered ($N_{\mathrm{det}}$), i.e.:
\begin{equation}
\label{eq:C2}
C_1 = c_0\, N_b\,\mathcal{J} \,N_{\mathrm{det}}\, \frac{\Tobs}{\Tfft}\,\mathcal{I}, 
\end{equation}
where $c_0\sim 6$~ms is a timed implementation cost per subpeakmap. 
We note that $\Tobs/\Tfft$ is exactly the number of FFT $\mathcal{N}$ only for gapeless data; $\mathcal{I}$ and $\mathcal{J}$ are factors equal to 2 for interlaced (by half) FFTs and frequency bands, respectively, and 1 otherwise. 

In other words, $C_1$ is the time required to perform steps from 2 to 6 listed in Sec.~\ref{Sec:method}.
The remaining step 7, needed to provide parameter estimates, is evaluated in a very cheap amount of time expressed by 
\begin{equation}
\label{eq:C2}
C_2 = c_{\theta} \, N_b^{'}\,\mathcal{J}, 
\end{equation}
where $c_{\theta}\sim 2.34$~s is an implementation and hardware-dependent computing cost per band. We note that $N_b^{'} \leqq N_b$ is the number of frequency bands where a signal has been identified.

Hence, for $\Tobs=1$~month, $\Tfft=512$~s, $N_b=131$, $\mathcal{J}=1$, $N_{\mathrm{det}}=1$, $\mathcal{I}=2$, and $N_{b}^{'}=128$, we have $C_1 \sim 2.3$~hours and $C_2 \sim 5$~minutes.

We underline that the search frequency is relevant to computational performance only because it affects the Doppler shift due to orbital motion. In fact, an increase in frequency would correspond to a reduction in $\Tfft$ [see Eq.~\eqref{eq:FFTlen}], and a consequent increase in $C_1$. However, since the presented novel method is extremely cheap from a computational point of view, even a relevant increase in computing time would keep the method computationally tractable\footnote{A similar reasoning applies when we reduce the source orbital period and increase the semi-major axis with respect to the choices applied here.}.

%****************************************************************************************************
\section{\label{Sec:Conc} Conclusions}
%****************************************************************************************************
This paper describes an incoherent and highly computationally cheap innovative method to search for continuous gravitational waves emitted by pulsars orbiting a companion object.
To show the pipeline performance, we analyse one month of simulated gapeless Gaussian noise single-detector data to which we added 131 CW signals emitted by pulsars in low- and high-eccentricity binary systems. We used an advanced LIGO-Virgo detector design sensitivity of $S_h = 4 \times 10^{-24}$~Hz$^{-1/2}$~\cite{AdvVirgoRef:2009,ADvLIGO}, and reported 128 detections, with the weakest -injected and detected- gravitational-wave amplitude of $h_0 \sim 10^{-24}$. We point out that, by using one year of data and three detectors in their advanced configuration, this translates into being able to detect CW signals with strain amplitude as low as $h_0 \sim 3 \times10^{-25}$. 
At very small frequencies this corresponds to reach, and go below, the torque balance limit currently foreseen for Scorpius~X-1~\cite{Leaci:2015bka}. 
We also stress that, as discussed in Sec.~\ref{Sec:SensEstim}, depending on the orbital Doppler modulation, our novel algorithm has the ability to detect also signals as low as $h_0 \sim 7 \times10^{-25}$ for a single-detector one-month data. So, the reachable sensitivity is expected to improve further with one year of three-detector data.

After claiming a detection, we recover the signal parameters, which we infer with decent accuracy. Note that, contrary to the Scorpius~X-1 MDC~\cite{ScoX1:MDC1} and previous searches \cite{ScoX1directAasi:2014qak}, in addition to circular orbits, we consider also high-eccentricity orbits, being able to process data at a strikingly cheaper computing time, and attaining a sensitivity comparable to some of the algorithms competing in \cite{ScoX1:MDC1} and to the technique presented in~\cite{Suvorova:2016rdc}. 

The current search may also be used as a fast quick-look analysis to scan the data, and possibly single out significant candidate signals which deserve further investigations.
The recovery of source parameters can be honed by assuming the estimates obtained for signal frequency and orbital parameters as preliminary. Hence, one can envisage to use those estimates to approximately demodulate the data from the binary orbital motion. This will serve to create longer FFTs, enhancing thus the search sensitivity. Rather than considering a fixed FFT duration for the entire search frequency range (as done here), we note that the sensitivity can be further increased by creating FFTs of different length in time, based on the highest frequency value of the subinterval into which the whole search frequency range has been previously split. This will allow a higher sensitivity at low frequencies.
At this point, the procedure outlined here can be reiterated to get more accurate signal parameter estimates.
Alternatively, or in addition, deep follow-up studies with a longer coherence time for the weakest detections can be applied to increase the SNR of a putative signal, and obtain even more precise parameter estimates (e.g. \cite{Shaltev:2014toa,NarrowBandVelaCrabAasi:2014jln}). This entails the generalisation of the re-sampling technique, currently used for CW signals from isolated pulsars~\cite{NarrowBandVelaCrabAasi:2014jln}, to the case of binary systems, and this is part of a separate ongoing study.

Although the current work is invaluable to both gauge and validate the performance of the various steps of the presented algorithm, accurate testing is needed (and currently underway) to enhance the Gaussian like filter rendering and efficiency. In fact, the success of such a filter varies depending both on the particular source orbital parameters, which govern the shape of the modulated signal pattern, and on the signal strength. 

Looking farther forward, we expect to generalise the method presented here to search for a broader class of signals, considering spindown parameters and accretion induced spin-wandering effects for the simulated sources, which will be added to real data collected from multiple detectors. 
We expect this novel methodology to be so robust not to be subject to frequency variability, and so the characteristics of spin wandering are expected to have no impact on the analysis. 
The complication to deal with real detector data will be faced applying \textit{ad hoc} noise reduction approaches that aim to identify, and possibly knock out, non-Gaussian artefacts, as well as coincidence-based approaches~\cite{NarrowBandVelaCrabAasi:2014jln}. %LeaciPhysSc2015

Furthermore, due to the tremendously cheap 2.4 CPU hours taken to analyse a fixed sky location, 131 1-Hz frequency bands, and one month of single-detector data, the current procedure can be applied in all-sky schemes in an equally successful and inexpensive way. This can be in fact achieved analyzing in parallel as many sky positions as permitted by the available computational power.

In addition, we plan to improve the current pipeline in order to provide estimates of source orientation and polarisation parameters (i.e., $\cosi$, $\psi$, and $\phi_0$), and also to get further enhancements in strain sensitivity, which will allow us to detect signal amplitudes at, or below, the torque balance limit currently foreseen for Scorpius~X-1-like sources. 

The present method -and its future improvements- will be applied to analyse new data collected by the ever-sensitive Advanced LIGO e Virgo detectors.
This is an oustanding challenge, which makes us more optimistic about being able to make direct detections of CW signals. Such detections will provide new insights into the internal structure, formation history and population statistics of neutron stars. In case of no detection, we plan however to set more astrophisically constraining upper limits on the gravitational-wave signal strength.

\section*{Acknowledgments}
The authors acknowledge the support of the University of Rome ``Sapienza'', Italian Istituto Nazionale di Fisica Nucleare (INFN), and the ``Rita Levi Montalcini Research Program".
This paper has been assigned document numbers LIGO-P1600233 and Virgo VIR-0365A-16.

%%%%%%%%%%%%%%%%%%%%%%%%%%%%%%%
\appendix

%****************************************************************************************************
\section{Maximal Doppler shift due to orbital motion}
\label{sec:maxim-doppl-shift}
%****************************************************************************************************

As shown in~\cite{Leaci:2015bka}, we estimate the maximal Doppler shift the intrinsic signal frequency of a
binary CW signal can undergo due to orbital motion. From the approximated phase-model of Eq.~\eqref{eq:ComplLinPhaseApprox} we
see that the instantaneous Doppler shift is
\begin{align}
  \label{eq:93}
  \left|\frac{\dot \phi(t)}{2\pi f}-1\right| &= \left|\frac{ \dot R(t)}{c}\right| \nonumber\\
  &= \asini\Om \left|\frac{ \sqrt{1-\ecc^2}\cos{E}\cos\argp - \sin{E} \sin\argp}{1 - \ecc\cos{E}}\right|\nonumber\\
  & \le \asini\Om \frac{|\sin{E}\sin\argp| + |\cos{E}\cos\argp|}{|1 -\ecc\cos{E}|}\,,
\end{align}
where we used the fact that $|a+b|\le|a|+|b|$ and $\sqrt{1-\ecc^2}\le1$.
In addition, we observe that
\begin{align}
  \label{eq:94}
  &|\cos{E}\cos\argp| + |\sin{E}\sin\argp| = \nonumber\\
  &\hspace*{0.3cm}\max\{|\cos(E+\argp)|,\, |\cos(E-\argp)|\} \le 1\,,
\end{align}
and $|1 - \ecc\cos{E}|\ge 1-\ecc$ to obtain
\begin{equation}
  \label{eq:MaxOrDS}
  \left|\frac{\dot \phi(t)}{2\pi f }-1\right|  =  \left|  \frac{\dot R(t)}{c} \right| \le \Delta M\,,
\end{equation}
with 
\begin{equation}
\label{eq:modDepth}
\Delta M = \frac{\asini\Om}{1 - \ecc}
\end{equation}
being the maximal Doppler modulation due to orbital motion.

%****************************************************************************************************
\section{Maximal FFT duration}
\label{Sec:Tsft-choice}
%****************************************************************************************************

The maximal length of the FFT is limited by the linear-phase approximation of Eq.~\eqref{eq:ComplPhase}. 
In order to improve sensitivity, we want to choose the longest
possible FFT duration $\Tfft$ with an acceptable error in the linear-phase approximation.
In order to estimate the maximal value of this phase-error ($|\Delta \phi|$) over a single FFT, we follow~\cite{Leaci:2015bka}
and estimate it as

\begin{equation}
\label{eq:FFTd}
|\Delta \phi| = \left|\frac1{2} \ddot \phi(t) \left(\frac{\Tfft}{2}\right)^2 \right|,
\end{equation}
but taking the second time derivative of the more precise phase of Eq.~\eqref{eq:ComplLinPhaseApprox} rather than that of Eq.~\eqref{eq:EccAnPhase}, as instead done in~\cite{Leaci:2015bka}. Hence, we have

\begin{equation}
\label{eq:Phiddot}
 \ddot \phi(t) = \frac{2\, \pi f \asini \Om^2 }{(1-e\,\cos E)^3} [ (\cos E - e) \sin \argp + \cos \argp \, \sin E \, \sqrt{1-e^2}],
\end{equation}
and replacing Eq.~\eqref{eq:Phiddot} in Eq.~\eqref{eq:FFTd} we obtain the following upper limit on the error in the linear-phase approximation: 

\begin{equation}
\label{eq:PhaseErrUL}
|\Delta \phi| \leq \frac{\pi}{4} \Tfft^2 \,  f \asini \Om^2 \frac{1+e}{(1-e)^3},
\end{equation}
where we used the fact that $|a+b|\le|a|+|b|$, $\sqrt{1-\ecc^2}\le1$, and 

\begin{align}
  \label{eq:compSinCos}
  &|\cos{E}\sin\argp| + |\cos \argp \sin E |= \nonumber\\
  &\hspace*{0.3cm}\max\{|\sin(E+\argp)|,\, |\sin(E-\argp)|\} \le 1\,,
\end{align}
which bring to

\begin{equation}
\label{eq:FFTlenGen}
\left| (\cos E -e )\sin \argp+ \cos \argp  \sin E \sqrt{1-e^2}  \right| \leq 1 + e.
\end{equation}
Hence, the FFT length must be 

\begin{equation}
\label{eq:FFTlen}
\Tfft \sim \frac{2 }{\Om} (1-e) \sqrt{\frac{1-e}{1+e} \, \frac{\Delta \phi}{\pi f \asini}}.
\end{equation}

%****************************************************************************************************
\section{Example of periodogram appearance}
\label{Sec:PerPlot}
\begin{figure}[htbp] 
  \raggedright (a)\hspace*{\columnwidth}\\[-0.9cm]
  \includegraphics[clip,width=\columnwidth]{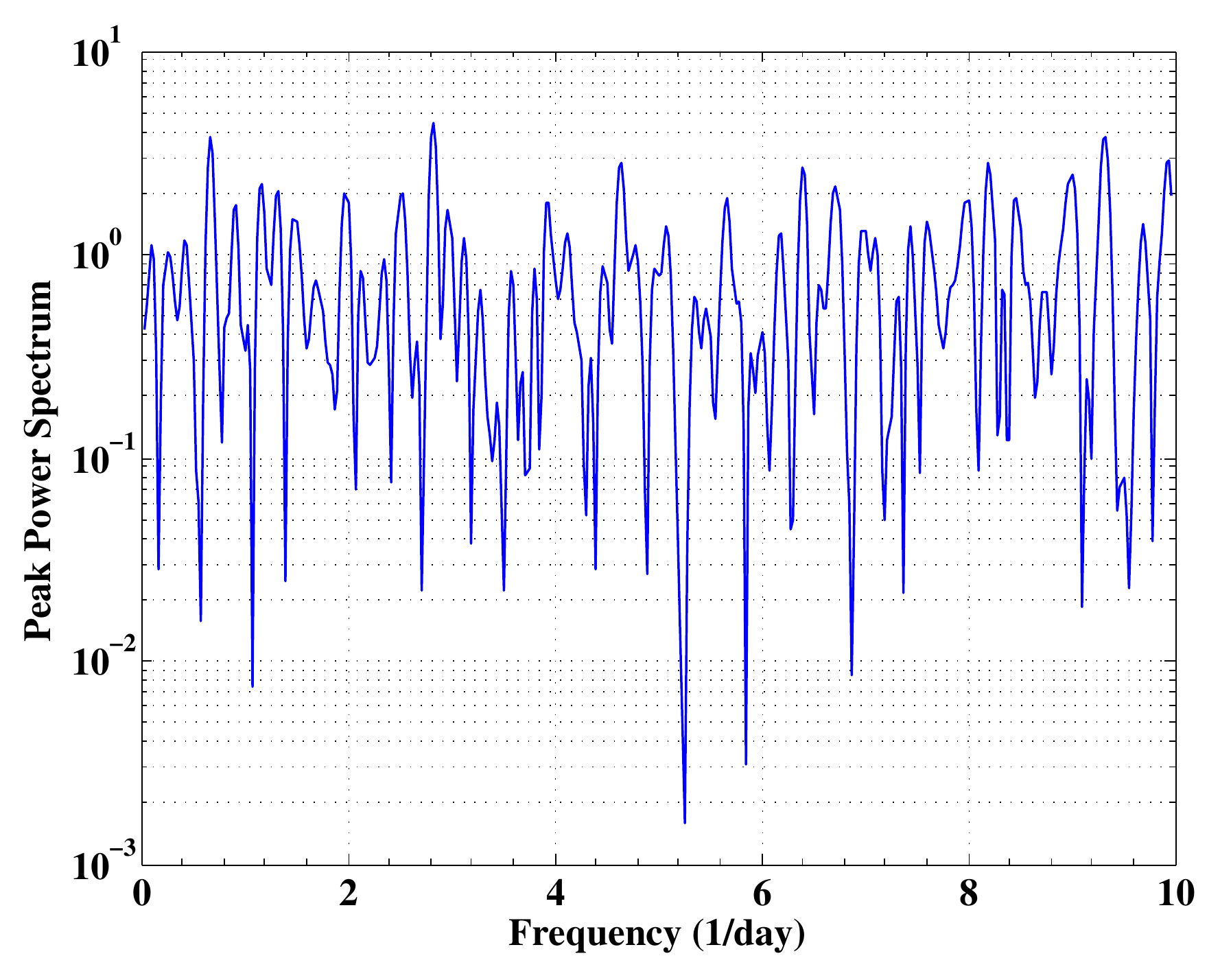} 
   \raggedright (b)\hspace*{\columnwidth}\\[-0.9cm]
  \includegraphics[clip,width=\columnwidth]{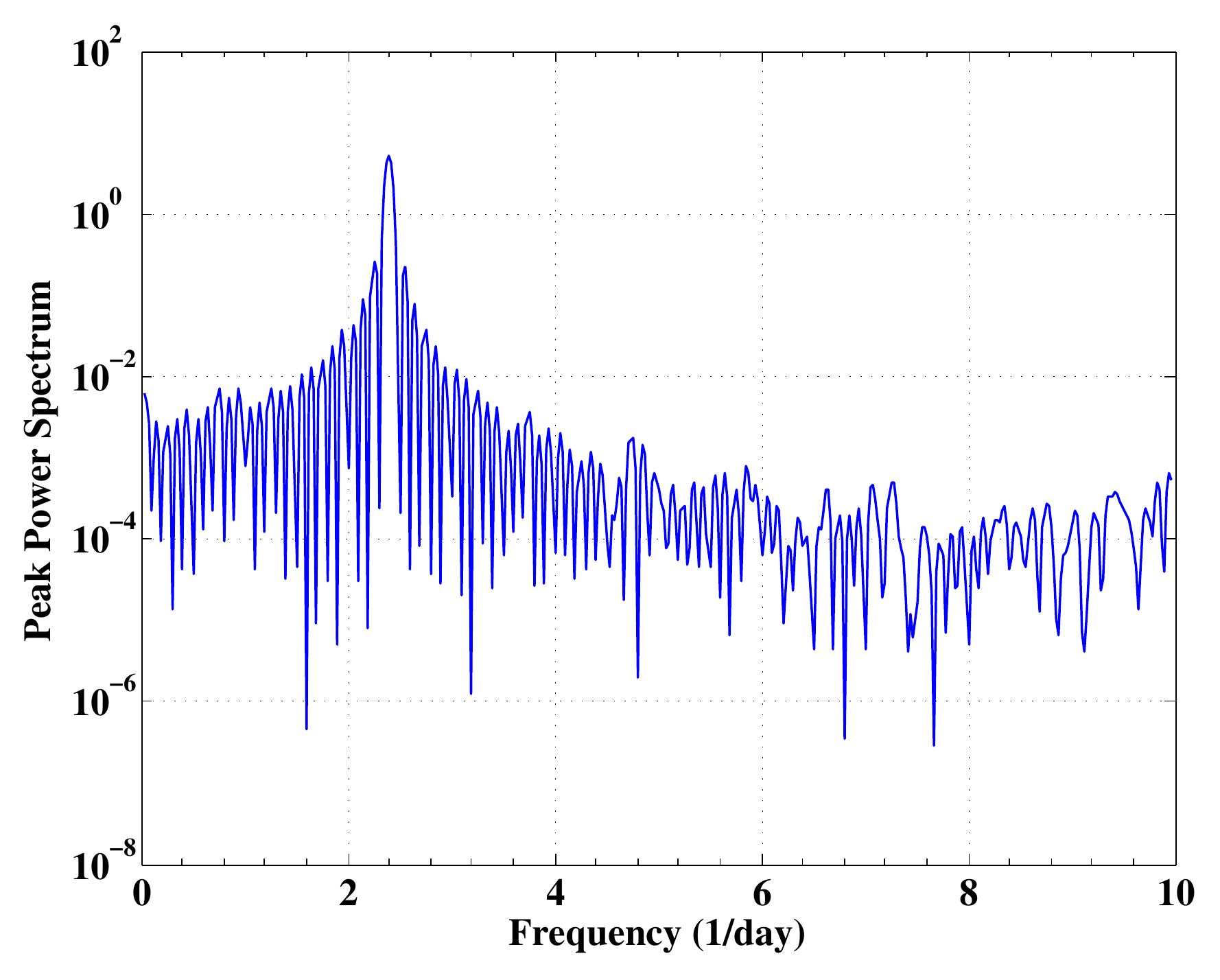}\\[-0.06cm]
  \raggedright (c)\hspace*{\columnwidth}\\[-0.9cm]
  \includegraphics[clip,width=\columnwidth]{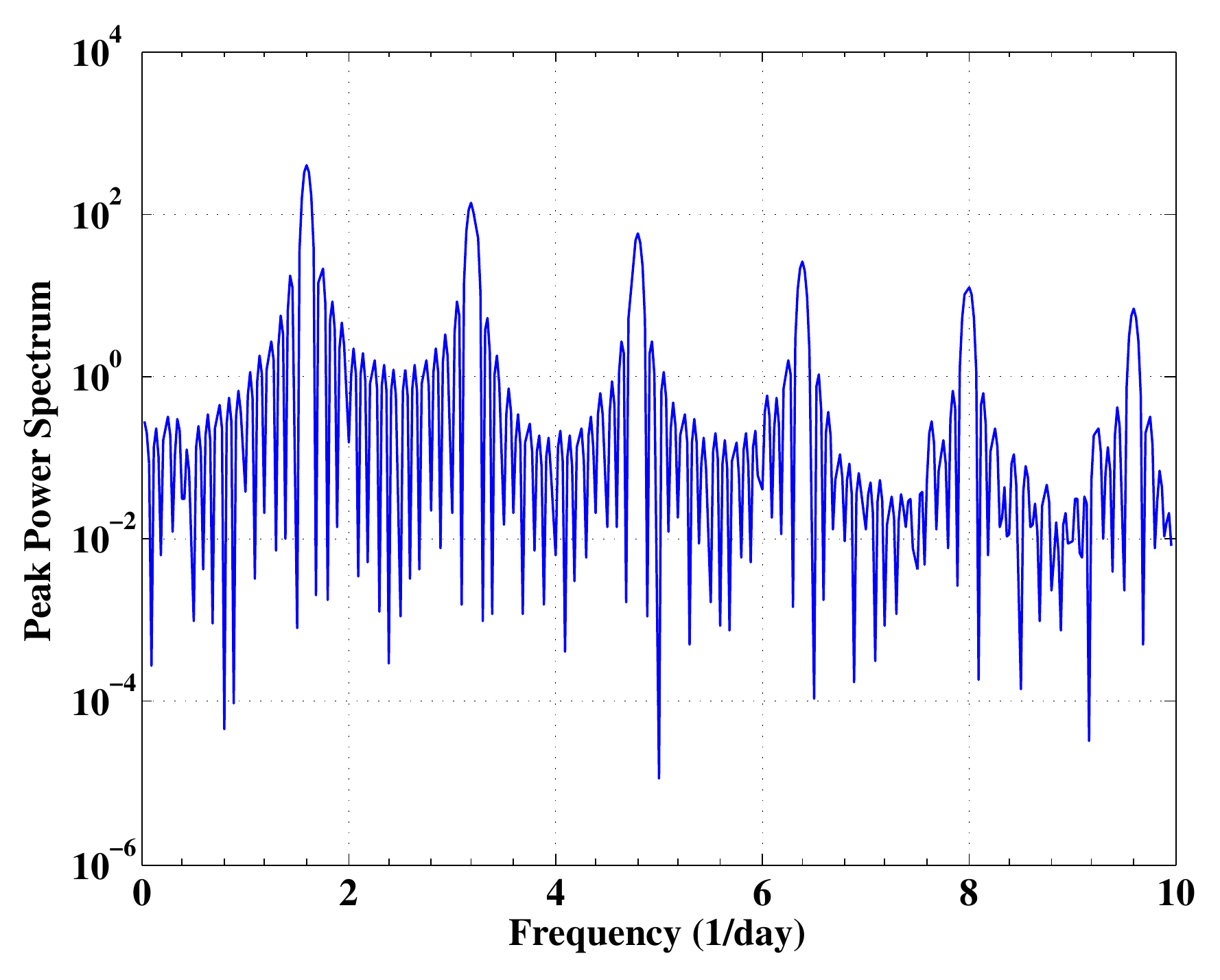}\\[-0.06cm]
  \caption{Semi-logarithmic plots of the periodogram as a function of the frequency for: normally distributed random numbers (a); set of synthetic data where two CW signals have been added at 155.5~Hz (b), and 1001.5~Hz (c), with orbital eccentricities $e=0$ (b), and $e\sim0.87$ (c), and strain amplitude $h_0\sim10^{-21}$ (b), and $h_0=4 \times 10^{-20}$ (c), respectively. The power spectral density used for these tests is $S_h = 4 \times 10^{-24}$~Hz$^{-1/2}$.
  The $x-$axis shows, for all panels, $400$ frequencies $\nu_j$ at which the periodogram of Eq.~\eqref{eq:peak_sp} has been evaluated for $\Tobs=10$~days and $r_\nu=4$.}
 \label{Fig:periodgr} 
\end{figure}

%**************************************************************************************************** 
Figure~\ref{Fig:periodgr} shows the periodograms evaluated for ($t_{c},\,\bar{f}_{p}$) pairs selected from randomly generated pure Gaussian noise [panel~(a)], and from data where two CW signals have been added into artificial gapeless Gaussian noise, with orbital eccentricity $e=0$ and $e\sim0.87$ for the panels (b) and (c), respectively. The artificially generated data sets span a period of $\Tobs=10$~days, and the conservative FFT duration used is $\Tfft = 128$~s. 

A strong component at $\nu \sim 2.4$~d$^{-1}$ is clearly observed in the periodogram of Fig.~\ref{Fig:periodgr}~(b), indicating the presence of a signal with orbital period $P = \nu^{-1} \sim 10$~h. The signal frequency is $\sim155.5$~Hz.   
In Fig.~\ref{Fig:periodgr}~(c) we can appreciate the harmonics due to the orbital eccentricity, which is $e \sim 0.87$. The first harmonic at  $\nu_1 \sim 1.6$~d$^{-1}$ is the fundamental, and the other harmonics are its multiples. All the harmonics are separated in frequency by $1/P$. The recovered orbital period is $P = \nu_1^{-1} \sim 15$~h. This simulated signal has a frequency of $1001.5$~Hz. 

We emphasise that both the power spectrum estimate, given by Eq.~\eqref{eq:peak_sp}, and the evaluation of the Lomb-Scargle periodogram bring to the same results. 

%****************************************************************************************************
\section{\label{spC} Harmonic content of the orbital Doppler modulation} 
%****************************************************************************************************
We estimate the harmonic content of the orbital Doppler modulation $\dot{R}(t)$ (i.e., the amplitudes of the fundamental and its multiples)
in order to understand what are the harmonics that contribute to estimate the orbital eccentricity. 

The time derivative of the R\o{}mer delay in Eq.~\eqref{RoemD} is given by
\begin{align}
  \label{OrDoppMod}
  &\frac{\dot{R}(t)}{c} = \asini \, \dot{E}(t)\,  \times \nonumber \\
  & \left[ -\sin \argp \, \sin E(t) + \cos \argp \, \cos E(t) \, \sqrt{1-\ecc^2}    \right]\,,
\end{align}
where $\dot{E}(t)=\Om / [1-\ecc\,\cos E(t)]$ is obtained deriving Eq.~\eqref{eq:KeplEq} with respect to time.

\begin{figure*}
\centering
\includegraphics[width=1.1\linewidth]{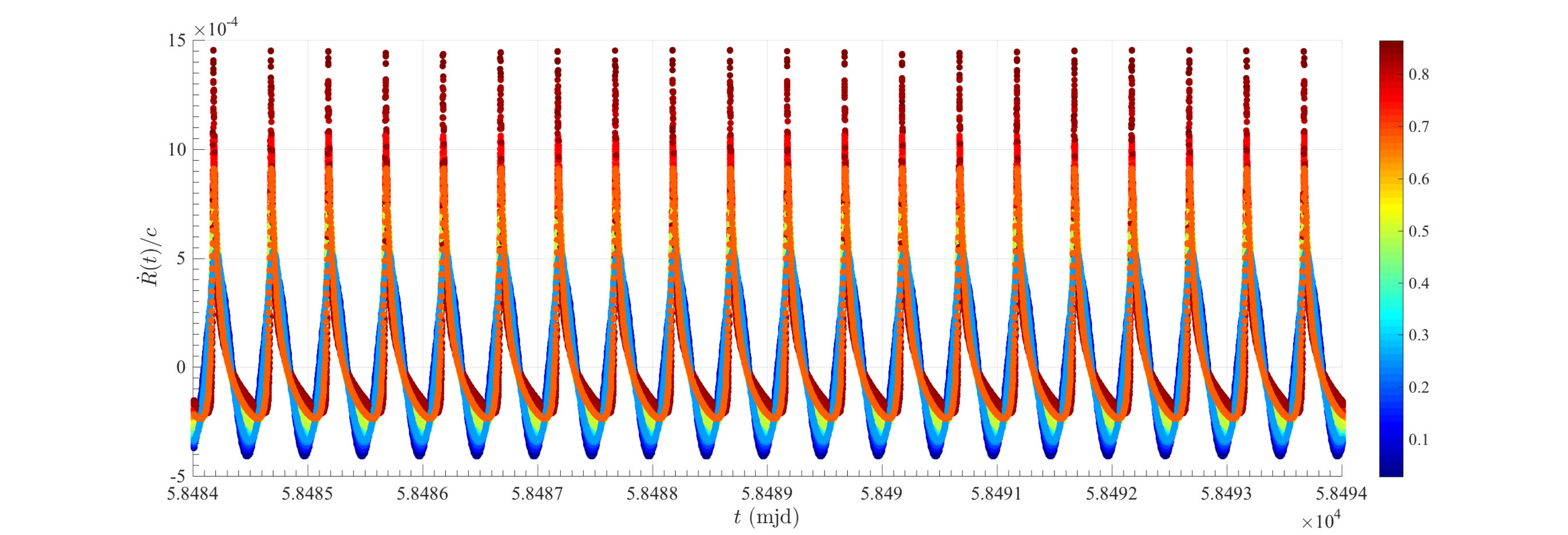} 
\includegraphics[width=1.1\linewidth]{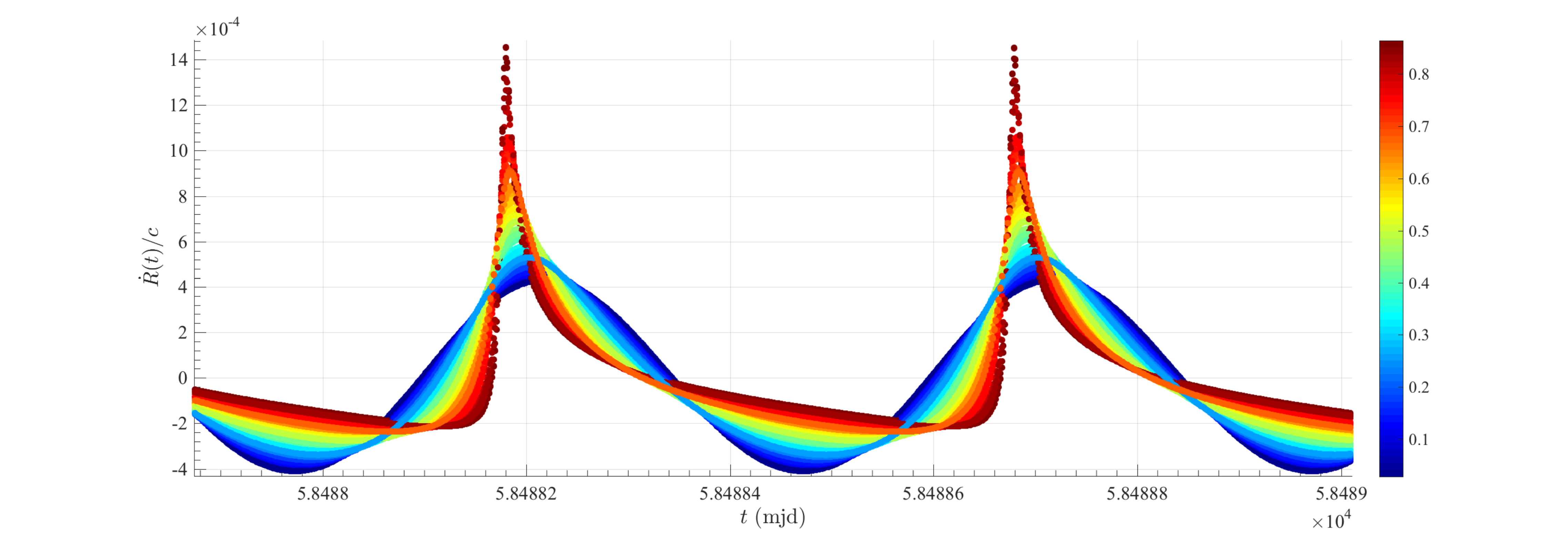}
\caption{\emph{Top panel}  Time derivative of the R\o{}mer delay, $\dot{R}(t)/c$ [given by Eq.~\eqref{OrDoppMod}], versus time for 60 simulated sources with fixed values of $\asini \sim 3$~s, $P~\sim 0.5$~d, $\argp \sim 6$~rad, $\tPeri \sim 54131$~mjd, but $\ecc$ randomly drawn from a uniform distribution in 
the range $\ecc\in[10^{-5},0.9]$, indicated by the values on the colour bar. \emph{Bottom panel} The same as top panel, but zooming into roughly a single period. \label{Fig:AndamRdotSc}}
\end{figure*}

In order to perform a spectral analysis of $\dot{R}(t)/c$, we first find the eccentric anomaly $E$ by numerically solving Eq.~\eqref{eq:KeplEq}
and inserting it in Eq.~\eqref{OrDoppMod}. We consider 60 fixed values of $\asini \sim 3$~s, $P~\sim 0.5$~d, $\argp \sim 6$~rad, $\tPeri \sim 54131$~mjd, and we draw $\ecc$ from a uniform distribution in 
the range $\ecc\in[10^{-5},0.9]$. Equation~\eqref{eq:KeplEq} is solved by employing standard iterative methods, such as the Newton's method and, in case of failure, the bisection method. 
In Fig.~\ref{Fig:AndamRdotSc} we plot the resulting $\dot{R}(t)/c$ as a function of time. From the top panel we observe, as expected, that each peak is separated from the next by the chosen orbital period $P~\sim 0.5$~d, and that sharp-edge peaks correspond to high eccentricity values (as shown on the vertical colour bar).

\begin{figure}
  \centering
  \includegraphics[width=\columnwidth]{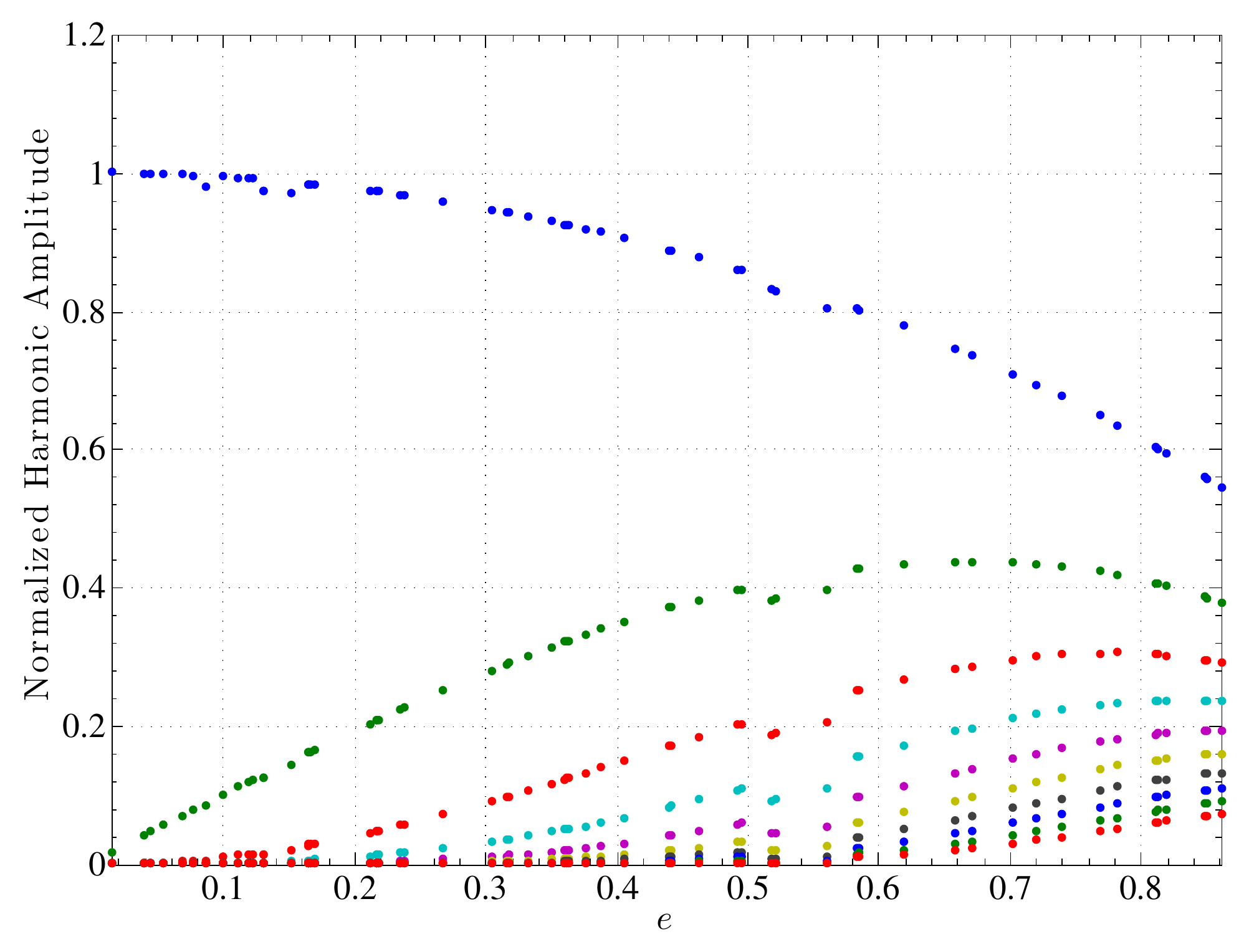} 
  \caption{(Color online) Normalised harmonic amplitude for the first 10 harmonics of $\dot{R}(t)/c$, and
  for 60 artificial sources with eccentricity $\ecc\in[10^{-5},0.9]$, and fixed $\asini, P, \argp, \tPeri$ (see main text for details). } 
 \label{fig:NormHarAmpEcc}
\end{figure}

Given the periodic nature of $\dot{R}(t)/c$ (see Fig.~\ref{Fig:AndamRdotSc}), we perform a Fourier decomposition of $\dot{R}(t)/c$ (by computing a discrete Fourier transform), and we plot in Fig.~\ref{fig:NormHarAmpEcc} the normalised harmonic amplitude for the first 10 harmonics
of $\dot{R}(t)/c$ as a function of the eccentricity, and for 60 values of $ \{\asini,\,\tPeri,\,P,\, e,\,\argp \}$. We stress that considering more points would only thicken the curves shown in Fig.~\ref{fig:NormHarAmpEcc}, without adding further information.
Independently of the chosen values for $\asini, P, \argp, \tPeri$, we note that, in order to estimate the
orbital eccentricity $\ecc$, only the first two harmonics of $\dot{R}(t)/c$ are necessary, whose contribution is
the most dominant. 

We further note that for very low-eccentricity orbits, only the first harmonic exists, which is not enough to precisely recover eccentricity values, which cannot then be distinguishable from zero, as stated in Sec.~\ref{RecoveryEstim}. This is confirmed by a separate investigation of the very low range $\ecc\in[10^{-8},10^{-3}]$.

From Fig.~\ref{fig:NormHarAmpEcc} we observe that the higher the eccentricity, the greater the number of contributing harmonics.

Furthermore, the amplitude of higher ($\geq10$) order harmonics is low enough that they can be entirely ignored.

%****************************************************************************************************
\section{Establishing existence of peak power spectrum harmonics}
\label{Sec:SecHarmPeakSpec}
%****************************************************************************************************
The peak power spectrum described in Sec.~\ref{PPS} is used to derive the source orbital period, which is given by the reciprocal of the fundamental frequency ${\nu}^m$, i.e. the first harmonic [see Eq.~\eqref{eq:orbitalP}]. The second harmonic (i.e. $2\,{\nu}^m$), when exists, is instead used to determine the source orbital eccentricity and argument of periapse [see Eqs.~\eqref{eq:eccRec} and~\eqref{eq:PerRec}, respectively]. In order to check if these harmonics exist in the peak power spectrum, we use a threshold based on a robust estimator, such as the median. 
After identifying ${\nu}^m$ and $2\,{\nu}^m$ in the peak power spectrum, we compute the median and dispersion parameter for roughly 80 samples around those two frequencies, and verify if the amplitude of each harmonic satisfy the following condition:

\begin{equation}
\label{eq:H12}
\mathcal{H}_h > \mathcal{M}(1) + \mathcal{M}(2), 
\end{equation}
with $h=1,2$, $\mathcal{M}(1)=\mathrm{median}(\mathcal{S}(h\,{\nu}^m - 40\,d\nu:h\,{\nu}^m + 40\,d\nu))$, and $\mathcal{M}(2)=\mathrm{median}( |\mathcal{S}(h\,{\nu}^m - 40\,d\nu:h\,{\nu}^m + 40\,d\nu) - \mathcal{M}(1) | )$.

%****************************************
\bibliography{DopplerModTrackP}
%****************************************

\end{document}